\documentclass[aps,prd,reprint,superscriptaddress,longbibliography,showpacs]{revtex4-2}

\usepackage{graphicx}% Include figure files
\usepackage{dcolumn}% Align table columns on decimal point
\usepackage{bm}% bold math
\usepackage[figurename=Figure]{caption}
\usepackage{ragged2e}
\DeclareCaptionJustification{plain}{\justifying}
\usepackage{float}    % For tables and other floats
\usepackage{verbatim} % For comments and other
\usepackage{amsmath}  % For math
\usepackage{bbold}
\usepackage{upgreek} 
\usepackage{amssymb}  % For more math
\usepackage{svg}
\usepackage{enumitem} % useful for itemization
\usepackage{latexsym,epsfig}
\usepackage{subcaption}
\usepackage{stackrel}
\usepackage{mathtools}
\usepackage{multirow}

\usepackage{braket}
\usepackage{placeins} %This contains \FloatBarrier command.

\usepackage[colorlinks=true,breaklinks=true,allcolors=blue]{hyperref}
\usepackage{lipsum}
\usepackage{dsfont}
\def\tpage{\ensuremath{{t_{\text{Page }}}}}

\usepackage[scr=boondox]{mathalfa}
\def\Hh{\ensuremath{{\cal H}}} %the Hamiltonian

\def\Ff{\ensuremath{{\cal F}}}

\def\Zz{\ensuremath{{\cal Z}}}

\usepackage[scr=boondox]{mathalfa}

\def\Tr{\ensuremath{{\operatorname{Tr}}}}

\newcommand\rj[1]{ {\color{magenta} #1} } %Rishabh's color is magenta

\usepackage{ulem}

\newcommand{\be}{\begin{equation}}
	\newcommand{\ee}{\end{equation}}
\newcommand{\bea}{\begin{equation}\begin{aligned}}
		\newcommand{\eea}{\end{aligned}\end{equation}}
\newcommand{\ben}{\begin{enumerate}}
	\newcommand{\een}{\end{enumerate}}

\newcommand{\entham}{entanglement Hamiltonian }
\newcommand{\renyi}{R{\'e}nyi }
\newcommand{\ment}{min-entropy }

\DeclareDocumentCommand{\nint}{ O{} O{} m }{\ensuremath{ \int_{\mbox{\scriptsize $#1$}}^{\mbox{\scriptsize$#2$}}\!\!\! \mbox{\small $\,\mathrm{d}#3$\! }}}

\usepackage{tikz,bm} % Useful for drawing plots
\usepackage{circuitikz}
\usetikzlibrary{decorations.markings}
\usetikzlibrary{decorations.pathreplacing,calligraphy}

\definecolor{mycolor}{rgb}{1,0.2,0.3}
\definecolor{brightgreen}{rgb}{0.4, 1.0, 0.0}
\definecolor{britishracinggreen}{rgb}{0.0, 0.26, 0.15}
\definecolor{cadmiumgreen}{rgb}{0.0, 0.42, 0.24}
\definecolor{ceruleanblue}{rgb}{0.16, 0.32, 0.75}
\definecolor{darkelectricblue}{rgb}{0.33, 0.41, 0.47}
\definecolor{darkpowderblue}{rgb}{0.0, 0.2, 0.6}
\definecolor{darktangerine}{rgb}{1.0, 0.66, 0.07}
\definecolor{emerald}{rgb}{0.31, 0.78, 0.47}
\definecolor{palatinatepurple}{rgb}{0.41, 0.16, 0.38}
\definecolor{pastelviolet}{rgb}{0.8, 0.6, 0.79}

\begin{document}
	
	\preprint{APS/123-QED}
	
	\title{Page Curve and Entanglement Dynamics in an Interacting Fermionic Chain}% Force line breaks with \\

	\author{Rishabh Jha}
	\email{rishabh.jha@uni-goettingen.de}
	\affiliation{%
		Institute for Theoretical Physics, Georg-August-Universit\"{a}t G\"{o}ttingen, Friedrich-Hund-Platz 1, 37077 G\"{o}ttingen, Germany
	}%
 \author{Salvatore R. Manmana}
	%\email{salvatore.manmana@theorie.physik.uni-goettingen.de}
	\affiliation{%
		Institute for Theoretical Physics, Georg-August-Universit\"{a}t G\"{o}ttingen, Friedrich-Hund-Platz 1, 37077 G\"{o}ttingen, Germany
	}%
 \author{Stefan Kehrein}
	%\email{Stefan.Kehrein@theorie.physik.uni-goettingen.de}
	\affiliation{%
		Institute for Theoretical Physics, Georg-August-Universit\"{a}t G\"{o}ttingen, Friedrich-Hund-Platz 1, 37077 G\"{o}ttingen, Germany
	}%


	\begin{abstract}

Generic non-equilibrium many-body systems display a linear growth of bipartite entanglement entropy in time, followed by a volume law saturation.
In stark contrast, the Page curve dynamics of black hole physics shows that the entropy peaks at the Page time $t_{\text{Page}}$ and then decreases to zero.
Here, we investigate such Page-like behavior of the von Neumann entropy in a model of strongly correlated spinless fermions in a typical system-environment setup, and characterize the properties of the Page curve dynamics in the presence of interactions using numerically exact matrix product states methods.
The two phases of growth, namely the linear growth and the bending down, are shown to be separated by a non-analyticity in the min-entropy before $t_{\text{Page}}$, which separates two different quantum phases, realized as the respective ground states of the corresponding entanglement (or equivalently, modular) Hamiltonian. We confirm and generalize, by introducing interactions, the findings of \href{https://journals.aps.org/prb/abstract/10.1103/PhysRevB.109.224308}{Phys. Rev. B 109, 224308 (2024)} for a free spinless fermionic chain where the corresponding entanglement Hamiltonian undergoes a quantum phase transition at the point of non-analyticity. 
However, in the presence of interactions, a scaling analysis gives a non-zero critical time for the non-analyticity in the thermodynamic limit only for weak to intermediate interaction strengths, while the dynamics leading to the non-analyticity becomes \textit{instantaneous} for interactions large enough. 
We present a physical picture explaining these findings. 
	\end{abstract}
	
	\maketitle
	
	%\tableofcontents

	%\section{Introduction}
	%\label{introduction section}
	
	%\phantomsection\label{sec:introduction}
	%\addcontentsline{toc}{section}{Introduction}
	
\section{Introduction}
\label{sec. introduction}

Quantum entanglement is the defining feature of quantum physics that differentiates it from classical physics. Fundamentally speaking, entanglement in a system implies the existence of a global state that cannot be separated as product states. Entanglement has gone through a journey from being an abstract ``spooky at a distance'' quantity to being a valuable resource used in quantum information and quantum computation \cite{Srivastava2024}. It has been used to study the properties of complex systems, such as in many-body physics, where the ground state properties help to characterize and classify the systems in consideration \cite{Amico2008May, Eisert2010Feb}. Recently, it has been realized that not just the equilibrium entanglement properties have universal features, but the entanglement dynamics also possess striking universal properties. For example, the temporal evolution of bipartite von Neumann entanglement entropy (vNEE) in non-equilibrium many-body systems has been used to differentiate ergodic systems from those with many-body localization \cite{Nandkishore2015Mar} when starting with a non-entangled initial state. In particular, many-body localized systems have logarithmic growth of the vNEE \cite{prosen, pollmann-mbl} while the many-body chaotic systems admit a linear growth \cite{chaos-1, chaos-2, chaos-3}. 
Entanglement as a resource has not just been analyzed in non-relativistic systems but studied in relativistic quantum field theories \cite{Witten2018Oct}, quantum fields in curved spacetime, and in cosmology \cite{Martin-Martinez2014Oct}. In fact, quantum entanglement has been proposed as a probe to the semiclassical structure of spacetime in quantum gravity \cite{Bianchi2014Oct} as well as a witness for quantum effects in gravity due to gravity-induced entanglement \cite{Marletto2017Dec}. One of the great paradoxes in modern physics revolve around the entanglement dynamics of black holes. A typical many-body system admits to a linear growth of the vNEE followed by a volume law saturation \cite{Calabrese2005Apr, Calabrese-lectures}. However, in case of black holes, a linear growth followed by a bending down of entropy (instead of saturation) happens which is famously known by the name of \textit{Page curve}.

\begin{figure}[t]
\centering
%\captionsetup{justification=centering}
\includegraphics[width=0.45\textwidth]{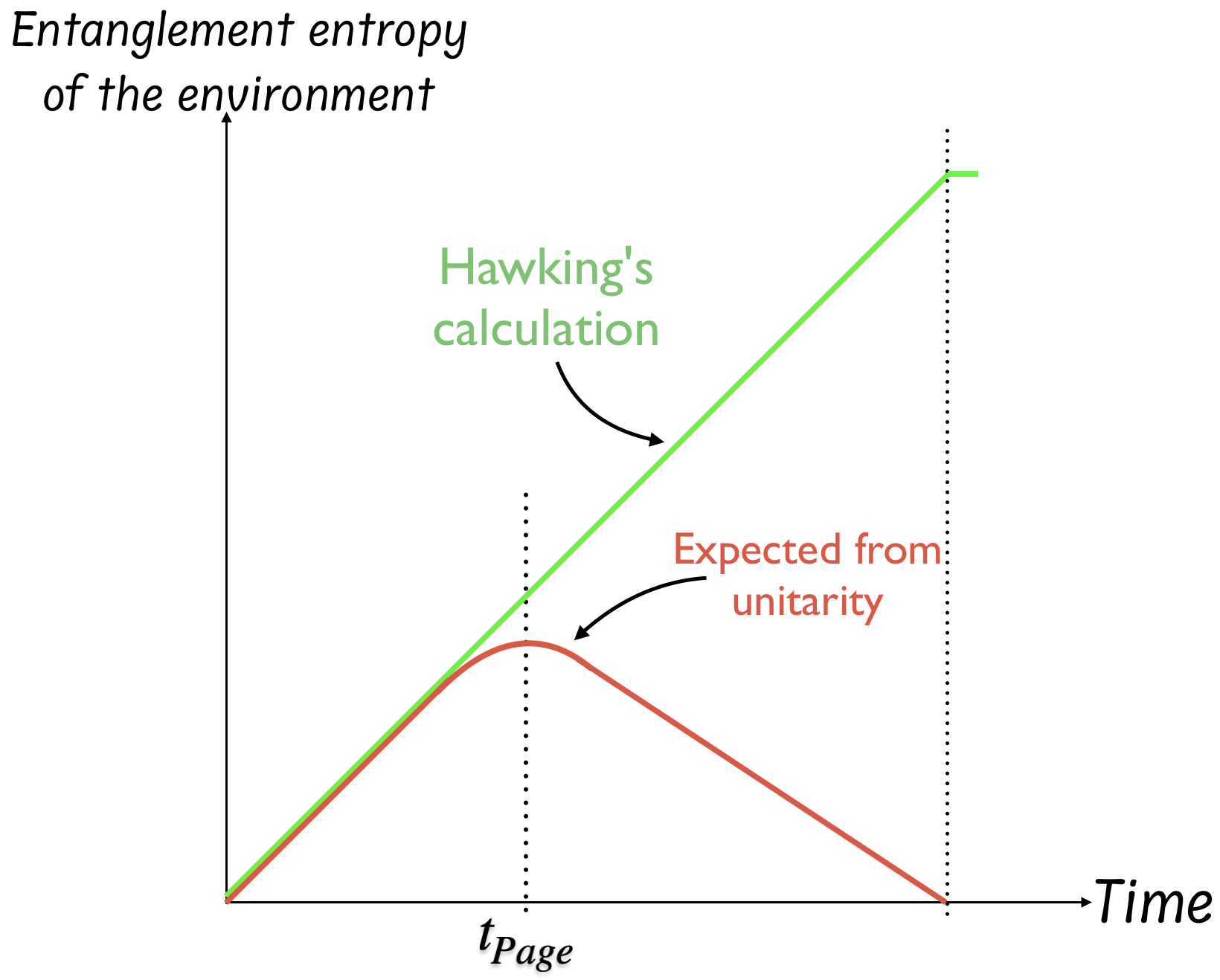}
    \caption{Illustration of the black hole information paradox: time evolution of the entanglement between the black hole and the Hawking radiation emitted (equivalent to the environment, quantified via the entanglement entropy) as expected from unitarity (Page curve, red), and as obtained from Hawking's semiclassical calculations (green).
    While unitarity predicts that the entanglement has to decrease after a certain time (called $\tpage$), the semiclassical calculation does not make such a prediction.
    %plotting entanglement entropy between the black hole and the radiation (equivalently the environment) with time where the expected curve from unitarity is the Page curve that bends down at $\tpage$, although Hawking's calculations show no such bending.
    }
    \label{page}
\end{figure}
The concept of Page curve has its origins in black hole thermodynamics which started with the seminal work of Bekenstein \cite{Bekenstein1972Aug, Bekenstein1973Apr, Bekenstein1974Jun} and Hawking \cite{Hawking1971May, Hawking1976Jan} who calculated the (\textit{coarse-grained}) entropy of a black hole and found it to be proportional to the area (instead of a volume). The revolution came when Hawking \cite{Hawking1974Mar, Hawking1975Aug} showed that black holes are endowed with temperature and, just like black bodies, emit thermal radiation precisely corresponding to this temperature. Thus, the four laws of black hole thermodynamics \cite{Bardeen1973Jun} were formulated analogous to standard thermodynamics. Suddenly the black holes were not really as dead as it was assumed to be. Hawking \cite{Hawking1974Mar} argued that if quantum theory is correct, then the black hole evaporation must be unitary, thereby conserving the information. To make this argument more precise, consider a universe with a black hole initially prepared in a pure state and an empty environment. As the black hole starts to evaporate, there exists radiation in the universe. So we can split the Hilbert space into the black hole region (labeled as the system) and the radiation (or equivalently the environment). Now if we start with a black hole in a pure state, then initially there is no radiation and the vNEE of each subsystem vanishes, i.e. $S(\Hh_{BH}) = S(\Hh_r) =0$. 
Finally, when the black hole has evaporated, we have an empty system that implies $S (\Hh_{BH}) = 0$, meaning that the entropy of the radiation (equivalently the environment) should \textit{self-purify} and we must have $S(\Hh_r) = 0$. Hence, there must be a bending down of entropy of the radiation at some point in time, such that the curve remains below the thermodynamic entropy of the black hole at all times \cite{Almheiri2021Jul}. But Hawking's calculation showed no such bending. This is illustrated in Fig.~\ref{page}. Hawking argued that there is a fundamental inconsistency between quantum physics and general relativity, and this has been dubbed as the \textit{black hole information paradox}. The phenomenon of bending down of the vNEE was first argued by Page \cite{Page1993Aug, Page1993Dec} who assumed conservation of unitarity which is a core principle of quantum physics. The expected curve from unitarity, as shown in Fig.~\ref{page}, is called the \textit{Page curve} where the maximum entropy reaches at the \textit{Page time} \tpage after which the entropy bends down. This happens because, as Page argued, the degrees of freedom inside the black hole that were entangled to the already emitted radiation are also evaporating, thereby reducing the vNEE between the bipartite system. Therefore, the bending of the vNEE curve is a crucial benchmark of any proposed resolution of the black hole information paradox. Page's argument fundamentally relies on the properties of the \textit{fine-grained} entropy, and it has been shown \cite{Almheiri2021Jul} that perturbative corrections to the Hawking state cannot reproduce the Page curve to all orders of perturbation. Therefore, if a solution exists, it has to be non-perturbative in the gravitational constant $G_N$.

Inspired by the black hole information paradox, we model a many-body system in this work where we have an arbitrary bi-partition which we refer to as the ``system'' and the ``environment''. We show that starting from a non-entangled initial state, we obtain a Page curve for the evolution of the vNEE. Analogous to black holes, the system is interacting while the environment is free. The situation where the system is also free is exactly solvable in the thermodynamic limit \cite{Kehrein2024Jun}. The free case has also been analyzed in Refs. \cite{Saha2024Dec, ganguly2025quantumtrajectoriespagecurveentanglement}. We also study \renyi entropies which all exhibit Page curve behavior where a recent study for local Hamiltonians for one dimensional chains found at most a linear in time behavior \cite{toniolo2025dynamicalalpharenyientropieslocal}. In particular the infinite order \renyi entropy (the so-called \textit{min-entropy}) develops a non-analyticity in its time evolution suggesting a breakdown of the semi-classical picture. Indeed, we find a quantum phase transition in the corresponding \entham where the critical time precedes the Page time. This allows us to draw a quantum phase diagram where the bending down of vNEE belongs to the quantum critical regime above the quantum critical point (see Fig.~\ref{fig:schematic for dynamical QPT}). We emphasize that the construction is \textit{analogous} to black hole physics but does not contribute directly towards the resolution of the black hole information paradox, albeit some parallels do exist between our construction and the one of black hole information paradox. 

The paper is organized as follows. We present the model in Section \ref{sec. model} where we also present the methodology and the results obtained in this work. We proceed to discuss the entanglement dynamics of the non-equilibrium case in Section \ref{sec. entanglement} where we explicitly study the entanglement entropy, min-entropy as well as the local particle number density for different parameter values.
We find a non-analytic behavior in the min-entropy, which we interpret (using the methodologies developed in Section \ref{sec. model} and Appendix \ref{subsec. entanglement hamiltonian}) as a quantum phase transition in the corresponding entanglement Hamiltonian. We provide a unified perspective on the time-dependent \entham for a model with conserved charge to explain the observed non-analyticity in Section \ref{sec. conserved charge picture}. In Section \ref{sec. exponent}, we perform finite-size scaling to extract universal finite-size scaling exponents across all interaction strengths. Near the non-analytic point (critical point), we determine the order parameter exponent $\beta$. Its value converges for all interaction strengths to that of the non-interacting limit in the thermodynamic limit, illustrating that the transition is a continuous phase transition and that the system belongs to the same universality class as the free counterpart. Finally, we extrapolate to the thermodynamic limit via performing a regression analysis in Section \ref{sec. regression} where we observe the persistence of non-analyticity at non-zero time for weak interactions in the system but vanishing for intermediate interactions. We also provide an analytical explanation and justification in Section \ref{sec. regression} for the observed behavior in the strongly interacting limit. We finally conclude in Section \ref{sec. conclusion}.
The appendices provide further details of the calculation and additional numerical results.

 \section{Model and Results}
 \label{sec. model}

\begin{figure}[t!]
    \centering
    %\captionsetup{justification=centering}
    \includegraphics[width=0.49\textwidth]{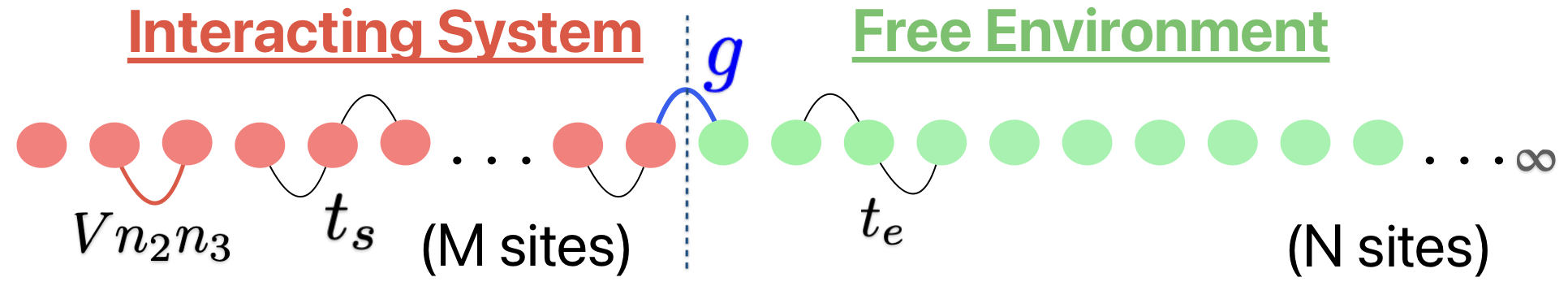}
    \caption{Sketch of the setup studied in this paper, see Eq.~\eqref{eq:model}: 
    The particles in the system ($M$ leftmost lattice sites, depicted in red) interact with each other via a nearest neighbor interaction $V$ (illustrated by the interaction term $Vn_2 n_3$ shown in red), and hop between neighboring sites with amplitude $t_s$. 
    The system is coupled to the environment ($N \gg M$ lattice sites on the right side, depicted in green) only by one hopping term of strength $g$ (depicted in blue). The particles in the environment hop between neighboring sites with amplitude $t_{e}$, but do not interact with each other. The total number of sites is $L=M+N$.}
    \label{setup}
\end{figure}

We are interested in a one-dimensional spinless fermionic chain with total number of sites $L=M+N$ whose Hamiltonian is given by 
    \begin{equation}
    \begin{aligned}
\Hh =& V  \left(\sum\limits_{i=1}^{M-1} n_i n_{i+1}\right) - t_s  \sum\limits_{i=1}^{M-1} \left( c_i^{\dagger} c_{i+1} + \text{h.c.}\right) \\
& - g \left( c_M^{\dagger} c_{M+1} + \text{h.c.}\right)  - t_{e} \sum\limits_{i=M+1}^{M+N-1} \left( c_i^{\dagger} c_{i+1}+ \text{h.c.}\right)
\label{eq:model}
\end{aligned}
\end{equation}
where $V$ is the interaction strength in the system containing $M$ sites, $t_s$ is the nearest-neighbor hopping strength within the system, $g$ is the tunneling strength connecting the system with the environment, and $t_{e}$ is the nearest-neighbor hopping strength within the environment consisting of $N$ sites. 
The operators $c_i^{(\dagger)}$ and $n_i^{\phantom{\dagger}} = c_i^{\dagger}c_i^{\phantom{\dagger}}$ are the usual fermionic annihilation (creation) operators and the particle density on lattice site $i$, respectively.
This setup has the structure
\begin{equation}
\begin{aligned}
    \Hh =&  \text{ (Interacting System) } + \text{ (Free Environment) } \\
    &+ \text{ (Coupling between the Sys. and the Env.) }
    \end{aligned}
\end{equation}
and is schematically shown in Fig.~\ref{setup}. If not stated otherwise, we set $t_s = t_e \equiv t_h \equiv 1$ throughout the paper.

We investigate the quench dynamics where the environment size is much larger than the system size ($N \gg M$) and hopping strengths are switched on at time $t=0.0$ where we start from the following initial state:
%\begin{widetext}
\begin{align}
| \text{Initial State} \rangle =& 
%| \text{Fully-Filled State in the System} \rangle_{\text{system}}  \nonumber \\ 
%&\otimes| \text{Empty Environment} \rangle_{\text{environment}} \, .
|\text{Fully-Filled}\rangle_{\text{sys}} \otimes |\text{Empty}\rangle_{\text{env}} \, .
\label{eq:initial state}
\end{align}
%\end{widetext}
Clearly the initial state is a product state of two pure states of the system and of the environment, respectively.
Therefore, the initial vNEE between the system and the environment is zero, and we are interested in studying the growth of entanglement dynamics post-quench.

In addition to the vNEE, we also study the \ment which is defined as the R{\'e}nyi entropy with infinite R{\'e}nyi parameter
\begin{equation}
    S_{\text{min}} \equiv \lim_{n \to \infty}S_n\left(\rho_{\text{sys}}\right)
\label{min-entropy definition}
\end{equation}
where $\rho_{\text{sys}}$ and $\rho_{\text{env}}$ are the reduced density matrices corresponding to the system and the environment (by tracing out the environment and the system, respectively) in Fig.~\ref{setup} and the \renyi entropy $S_n$ is defined as
\begin{equation}
S_n\left(\rho_{\text{sys}}\right)=\frac{1}{1-n} \ln \operatorname{Tr}_{\Hh_{\text{sys}}}\left(\rho_{\text{sys}}^n\right)=S_n\left(\rho_{\text{env}}\right)   .
\label{renyi entropy definition}
\end{equation} 
Note that $\lim_{n \to 1} S_n$ recovers the vNEE. The reason for studying the \ment is explained in Appendix \ref{subsec. entanglement hamiltonian} where we discuss the concept of \textit{entanglement Hamiltonian} $\Hh_E$ \cite{Dalmonte2022Nov}. 
Here we provide an interpretation \cite{Kehrein2024Jun} of min-entropy based on the distance from the closest separable state $|\Psi\rangle_{\text{separable}} = |\psi\rangle_{\text{sys}} \otimes |\phi\rangle_{\text{env}}$ such that the min-entropy is given by the negative of the logarithm of the maximum overlap of any state $\ket{\Phi(t)}$ with $\ket{\Psi}_{\text{separable}} $. Therefore, the smaller the value of min-entropy, the closer the state $|\Phi(t)\rangle$ is with the nearest separable state $|\Psi\rangle_{\text{separable}} $. 

In Appendix \ref{subsec. entanglement hamiltonian}, we introduce a related concept of \entham where we interpret the min-entropy ($\lim_{n \to \infty} S_n$) as the ground state eigenvalue of the entanglement Hamiltonian. The excited states correspond to a fictitious temperature $1/n$ where $n$ is the \renyi parameter (also see Fig.~\ref{fig:schematic for dynamical QPT}).

\subsection{Results}
\label{subsec. results}

\textit{Free case.---} The free version ($V=0$) of our model (Eq.~\eqref{eq:model}) has been exactly solved in the thermodynamic limit in Ref.~\cite{Kehrein2024Jun}. We summarize briefly the findings of the free case and proceed to present the results for the more general interacting case. The Peschel-Eisler formalism \cite{Peschel2009Dec} was employed for the free case to calculate the exact entanglement entropy from reduced density matrices. Since the free Hamiltonian is quadratic which can be diagonalized exactly, the analytical solutions in the weak-coupling limit between the system and the environment are possible to be obtained in the thermodynamic limit. This is done by mapping the free model to a set of $k-$disjoint resonant level models (RLMs) where universal behavior for the decay of impurity orbital occupation is used in the wide, flat band limit and the associated reduced density matrix of the system is obtained as direct product of reduced density matrices of resonant level modes for $k$-disjoint RLMs. Indeed, the vNEE follows a Page curve like dynamics and is shown to match against the numerical results (which has the limitation due to particle reflection at the boundary, something we will also face below for our interacting case). Furthermore, the Page curve behavior is obtained for all \renyi entropies. In particular, the min-entropy admits a non-analytical behavior at a critical time that precedes the Page time. This is demonstrated both analytically and numerically, and corresponds to a quantum phase transition in time of the corresponding \entham (as discussed above). The bending down of the vNEE is argued to belong to the quantum critical regime corresponding to this quantum critical point of non-analyticity, thereby leading to two different quantum phases of matter. This justifies the breakdown of the semi-classical picture relating the particle current escaping the system and the entanglement generation beyond the Page time \cite{Calabrese2005Apr}.

\textit{Our results.---} In this work, we consider a general interacting case as in Eq.~\eqref{eq:model} where we study the entanglement dynamics starting from a product state as in Eq.~\eqref{eq:initial state}. We show the persistence and robustness of the features obtained in the case of free model \cite{Kehrein2024Jun}. We use time-dependent Matrix Product State methods (MPS) \cite{Schollwoeck201196,Paeckel2019} implemented in the package SciPAL-SymMPS \cite{symmps} to calculate the local observables, the entanglement entropy as well as the largest four Schmidt coefficients from which the ground state (associated with min-entropy) and the first three excited states of the \entham are calculated. 
We impose an open boundary condition in our MPS calculations and go to $L=50$ lattice sites where we consider different system sizes, namely $M=3,4,5,6$ and $7$ (consequently the environment sizes are $N = L - M$) to be able to perform scaling analysis. In order to generate the initial state in Eq.~\eqref{eq:initial state} in our MPS calculations, we imposed a strong potential in the environment via the chemical potential ($\mu = 10^6$). Then the initial state (Eq.~\eqref{eq:initial state}) was used to start the time evolution using the 2-site time-dependent variational principle (TDVP) method \cite{tdvp-1, tdvp-2}, where the maximum bond dimension was set to $\chi = 5000$ with a discarded weight of $10^{-12}$.
The details of all parameters used in our MPS calculations can be found in the bash scripts used for the package SciPAL-SymMPS which we have made publicly available along with the data generated and used in this work \cite{zenodo}.
The main results are summarized as follows:
\begin{enumerate}[label=(\roman*)]
    \item We obtain a Page curve dynamics for the evolution of the vNEE in time. 
    We observe that as the interaction strength is increased, the maximum of the vNEE at the Page time is reduced, signifying lesser entanglement between the system and the environment. 
    \item The min-entropy shows a non-analyticity despite switching on the interaction, leading to a quantum phase transition in the associated \entham corresponding to a fictitious temperature of $1/n$. The persistence of the quantum phase transition in the presence of interactions illustrates the robustness of the two quantum phases of matter separated by the quantum critical point (non-analytic) in the ground state of the entanglement Hamiltonian. The fundamental reason is the conserved charge picture (Section \ref{sec. conserved charge picture}) where the \entham conserves the particle number, leading to different sectors of conserved charges crossing each other at the point of criticality as the particle gets emitted from the system into the environment. 
    \item We investigate the nature of the quantum phase transition through finite-size scaling. Figure \ref{fig:collapse plot V=1.2} demonstrates excellent data collapse across all interaction strengths (additional plots are provided in Appendix \ref{app. exponent}), confirming universal finite-size scaling exponents (Table \ref{table:exponents}). Furthermore, we find a universal order parameter exponent $\beta \sim 0.5$ (Table \ref{table:lambda exponent}) across all interaction strengths, confirming a continuous quantum phase transition. This value agrees with the thermodynamic limit result $\beta = 0.5$ for the non-interacting chain in Ref. \cite{Kehrein2024Jun} (Eq. (61) therein), suggesting the transition belongs to the same universality class.
    \item The slope of min-entropy is (essentially) linear until the non-analytic point, and this allows us to do a scaling analysis and extrapolate to the thermodynamic limit using linear regression. We find in the thermodynamic limit that for weak interaction ($V/t_h < 1.0$), there exists a finite, non-zero critical time at which the non-analyticity appears while for intermediate interactions ($V/t_h \sim 1.0$), the critical time tends to zero. We also find an acute sensitivity on the coupling strength $g$ between the system and the environment (Tables \ref{table:regression_critical_decay_fraction} and \ref{table:appendix_regression_critical_decay_fraction}) where a weaker coupling $g$ requires a stronger interaction for the thermodynamic critical value to vanish.
    \item We provide an interpretation for this phenomenon: when the first particle leaves the system, a hole enters the system from the environment. As energetically favorable process, the hole does not propagate further into the system and all the consequent holes entering the system develop a domain-wall structure as shown in Fig.~\ref{brick-wall}. Accordingly, in the strong interacting limit ($V/t_h \gg 1.0$), one particle decays into the environment and the system gets stuck with no further dynamics happening. This allows us to write an ansatz for the local number density in the system (Eq.~\eqref{eq:ansatz}). As such, we solve for the critical time where the eigenvalues of the \entham cross (a quantum phase transition occurs that is represented as the non-analyticity in min-entropy) for large values of $M$ (system size) and show that the interpretation fits the aforementioned result of critical time tending to zero for increasing interaction strengths in the thermodynamic limit. 
\end{enumerate}

 \section{Entanglement Dynamics and Page Curve}
\label{sec. entanglement}

As discussed above, we present the entanglement dynamics of \renyi entropies in this section for a total system size $L=50$ for model \eqref{eq:model}. Motivated by the black hole setup, in the following, we have $N \gg M$ while $L=M+N=50$ is fixed.
We consider different system sizes, namely $M=3,4,5,6$ and $7$ in order to perform a scaling analysis $M\to\infty$ to the thermodynamic limit (TL), see Section~\ref{sec. regression}. Since in all cases treated we consider time scales before any particles reach the right edge of the lattice at site $L$, in this way we indeed obtain the behavior in the TL of the entire set up.
The parameter values are fixed in the model Eq.~\eqref{eq:model} as $t_s = t_{e} = 1 \equiv t_h$ and the system and the environment are coupled weakly, in particular $g \in \{0.25, 0.50\}$. Two interaction regimes are considered: weak ($V/t_h < 1.0$) and intermediate ($V/t_h \sim 1.0$) whereas an analytical explanation is provided in Section \ref{sec. regression} for the strongly interacting case ($V/t_h \gg 1$).
\begin{figure}
\centering
%\captionsetup{justification=centering}
\includegraphics[width=0.49\textwidth]{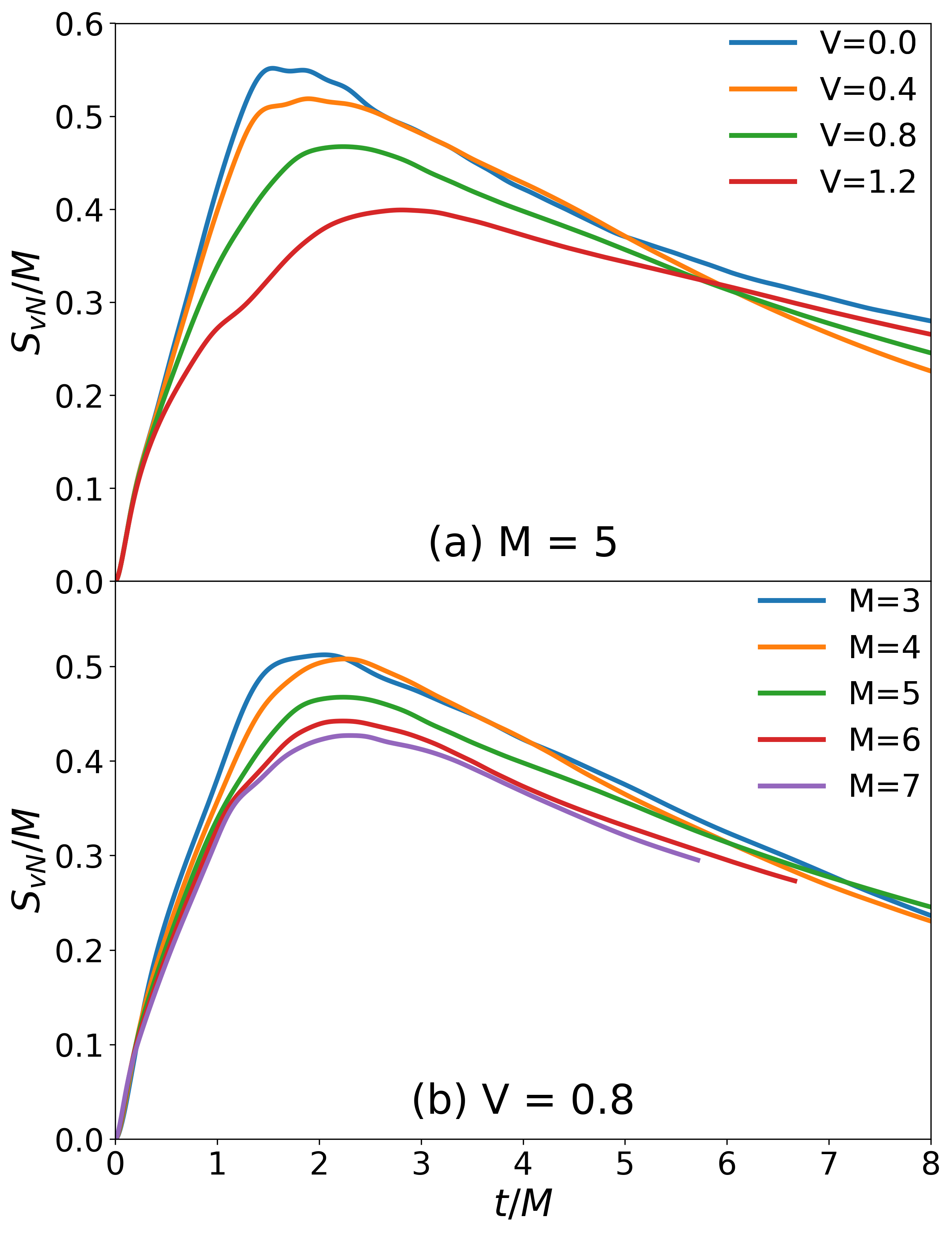}
    \caption{MPS results for the time evolution of the von Neumann bipartite entanglement entropy density [Eq.~\eqref{renyi entropy definition} for $n \to 1$] between the system and the environment for a chain with in total $L=50$ lattice sites, $t_s=t_e=1$, $g=0.5$, and interaction strength $V$ as indicated [see model \eqref{eq:model}].  
    Note that we scaled the time-axis by the total number of particles $M$, which in our setup is identical with the size of the system. %Total number of sites are fixed at $L=50$. 
    %The parameters in Eq.~\eqref{eq:model} are fixed at $t_s = t_e = 1.0$ and $g=0.5$. 
    The simulation is run for times starting at $t=0.0$ to a maximum time of $t_{\text{max}} = 40.0$ with a time step size $\mathrm{d}t = 0.02$. 
    (a) Behavior at fixed system size  $M=5$ for different values of $V$. 
    (b) Behavior at fixed $V=0.8$ for different system sizes $M=3, \, \ldots, \, 7$. 
    We find in all cases a robust Page curve where the peak value of the entanglement decreases as either $V$ or $M$ is increased. %(b) We fix the interaction strength at $V=0.8$ and plot for different system sizes where we find the Page curve behavior to be robust.
    }
    \label{fig:ee_in_time}
\end{figure}

\begin{figure}
\centering
%\captionsetup{justification=centering}
\includegraphics[width=0.49\textwidth]{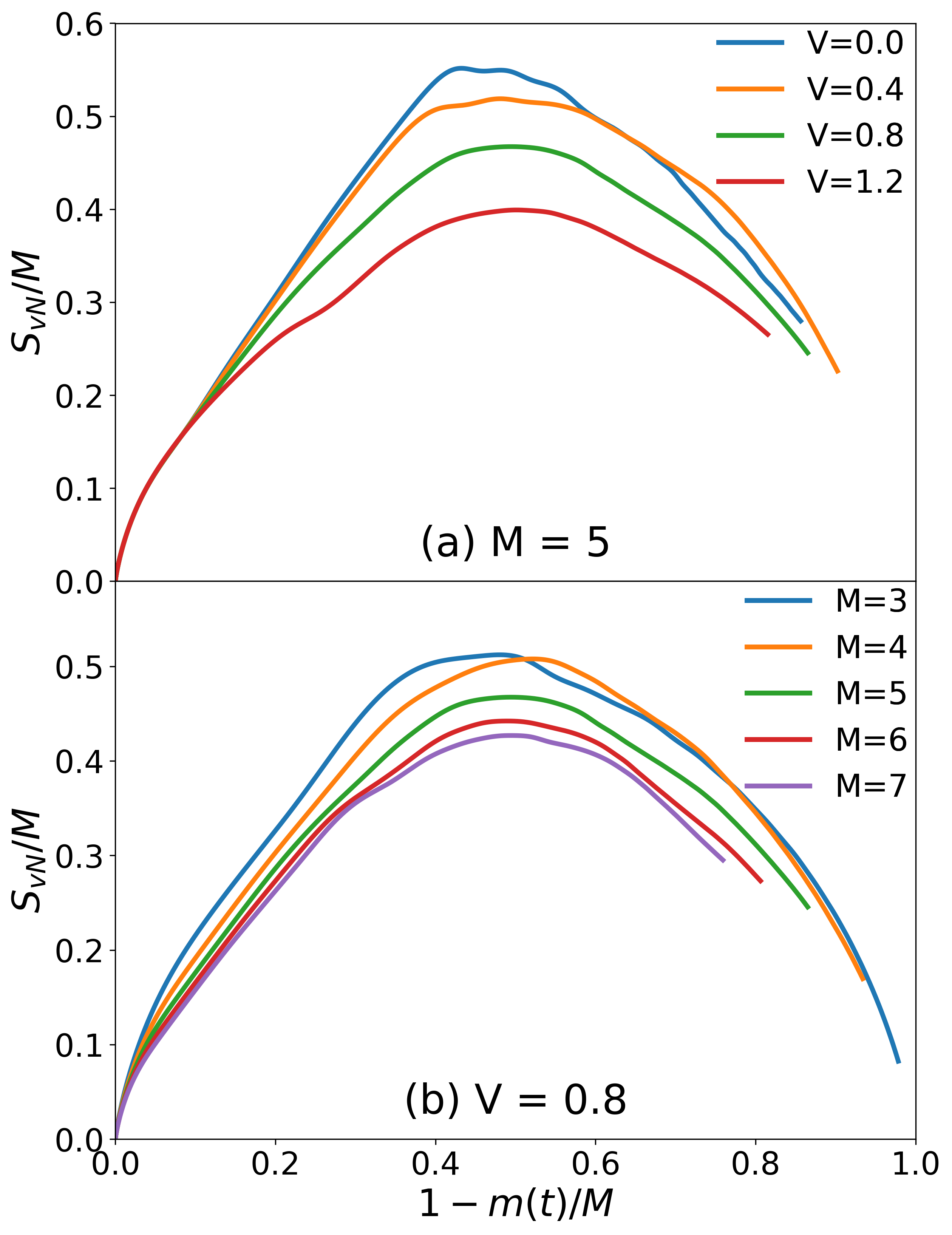}
    \caption{Same as in Fig.~\ref{fig:ee_in_time}, but plotted as a function of the  decayed fraction of particles $1-m(t)/M$, which escaped from the system into the environment.}
    \label{fig:ee}
\end{figure}

\textit{Page curve and effect of interaction.---} 
In Figs.~\ref{fig:ee_in_time} and \ref{fig:ee}, we present our results for the vNEE ($n=1$ \renyi entropy, in the following called $S_{\rm vN}$) plotted against the decayed fraction of particles from the system into the environment and the scaled time, respectively. We find the Page curve entanglement dynamics in time in Fig.~\ref{fig:ee_in_time}, analogous to the black hole case. For further detailed analysis, in Fig.~\ref{fig:ee}, we plot the vNEE density (=$S_{\text{vN}}/M$) against the fraction of particles decayed into the environment $=1 - m(t)/M$, where $m(t)$ is the number of particles in the system at time $t$. 
We clearly observe a growth up to a maximal value, which is reached at what we in the following call Page time $t_{\rm Page}$, and afterward a pronounced decay.
The maximum possible vNEE density is $\ln(2) \approx 0.69$ where we find that increasing the interaction in the system suppresses the maximum entanglement between the system and the environment at the Page time. The figure is plotted for the system size $M=5$ while plots for other system sizes are provided in the Appendix \ref{appendix}.

\begin{figure}
\centering
%\captionsetup{justification=centering}
\includegraphics[width=0.49\textwidth]{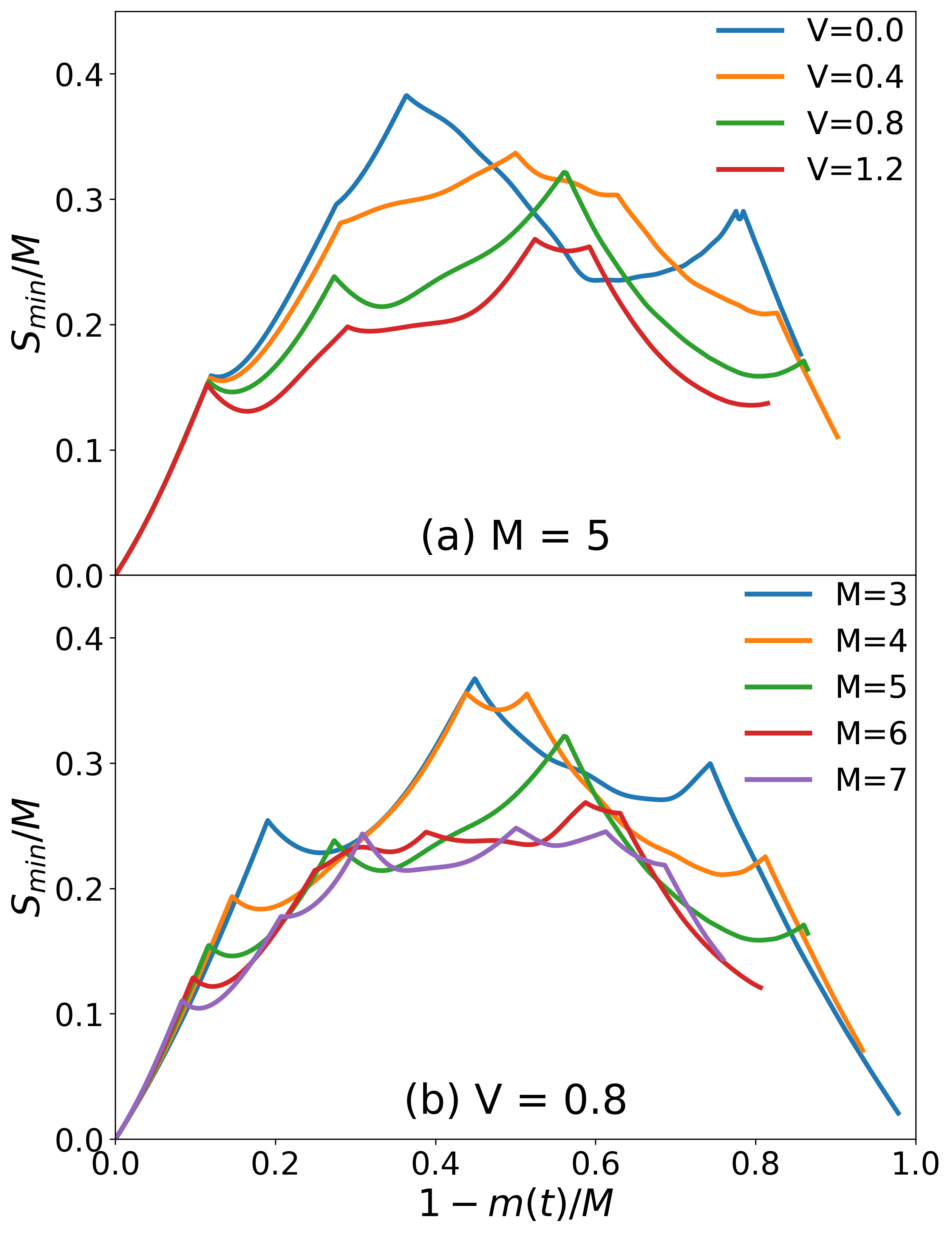}
    \caption{MPS results for the time evolution of the Min-entropy density [Eq.~\eqref{min-entropy definition}] plotted agains the decayed fraction $1-m(t)/M$ and for the same parameters as in Fig.~\ref{fig:ee}.  
    %Min-entropy density between the system and the environment is plotted against the decayed fraction of particles from the system into the environment, 
    Non-analyticities in the temporal evolution are observed in all cases. 
    %where we observe non-analyticity developed over temporal evolution. We study for the same parameters as we study the von Neumann entropy in Fig.~\ref{fig:ee}. 
    (a) For fixed system size at $M=5$ and different interaction strengths $V$, the first non-analyticity happens (to a very good approximation) at the same value of the decayed fraction for all values of $V$ shown. 
    %There is a robust behavior of non-analyticity that persists at the same time, independent of tuning the interaction strengths. As shown for the free case in Ref.~\cite{Kehrein2024Jun}, the later non-analyticities smear out in the thermodynamic limit. 
    (b) For fixed interaction strength $V=0.8$ and varying the system size $M$, we find that the first non-analyticity develops at earlier times as $M$ increases. 
    %Due to the linear behavior of min-entropy until the first non-analyticity, we perform a linear regression for the non-analytic points in Section \ref{sec. regression} to extrapolate to the thermodynamic limit.
    }
    \label{fig:min-ee}
\end{figure}

\textit{Non-analyticity in min-entropy.---} We study the min-entropy that is defined in Eq.~\eqref{min-entropy definition} and we find a non-analyticity in the temporal evolution at some critical time $t_c <$ \tpage. 
Fig.~\ref{fig:min-ee}(a) displays data for $M=5$ for different values of $V$; the behavior for further values of $M$ is presented in Appendix~\ref{appendix}. As can be seen, several non-analytical points appear as a function of time.
In the following, we mainly focus on the first non-analyticity, since it will help us in better understanding the behavior before reaching the maximum of the entanglement.
As can be seen in Fig.~\ref{fig:min-ee}(a), the value of $V$ has a minimal influence on the value of the decayed fraction $1-m(t)/M$, at which the first non-analyticity appears.
This illustrates the persistence of the non-analytic behavior also in the presence of interactions. The question arises, if the behavior is universal for all values of the interactions, or if eventually the interactions will lead to different behavior. To study this, we investigate the behavior in the TL. In Fig.~\ref{fig:min-ee}(b) we now keep $V$ fixed and study the dependence of the critical value of $1-m(t)/M$ on $M$.
We see that the kink position happens at smaller values of $1-m(t)/M$ when increasing $M$. 
This is further studied in Section~\ref{sec. regression}, where we extrapolate this value to the TL via linear regression.
We refrain from studying the later non-analyticities, since we expect that, similar to the findings in the free case~\cite{Kehrein2024Jun}, in the TL they will smear out, so that only the first non-analyticity will persist.

\begin{figure}
\centering
%\captionsetup{justification=centering}
\includegraphics[width=0.49\textwidth]{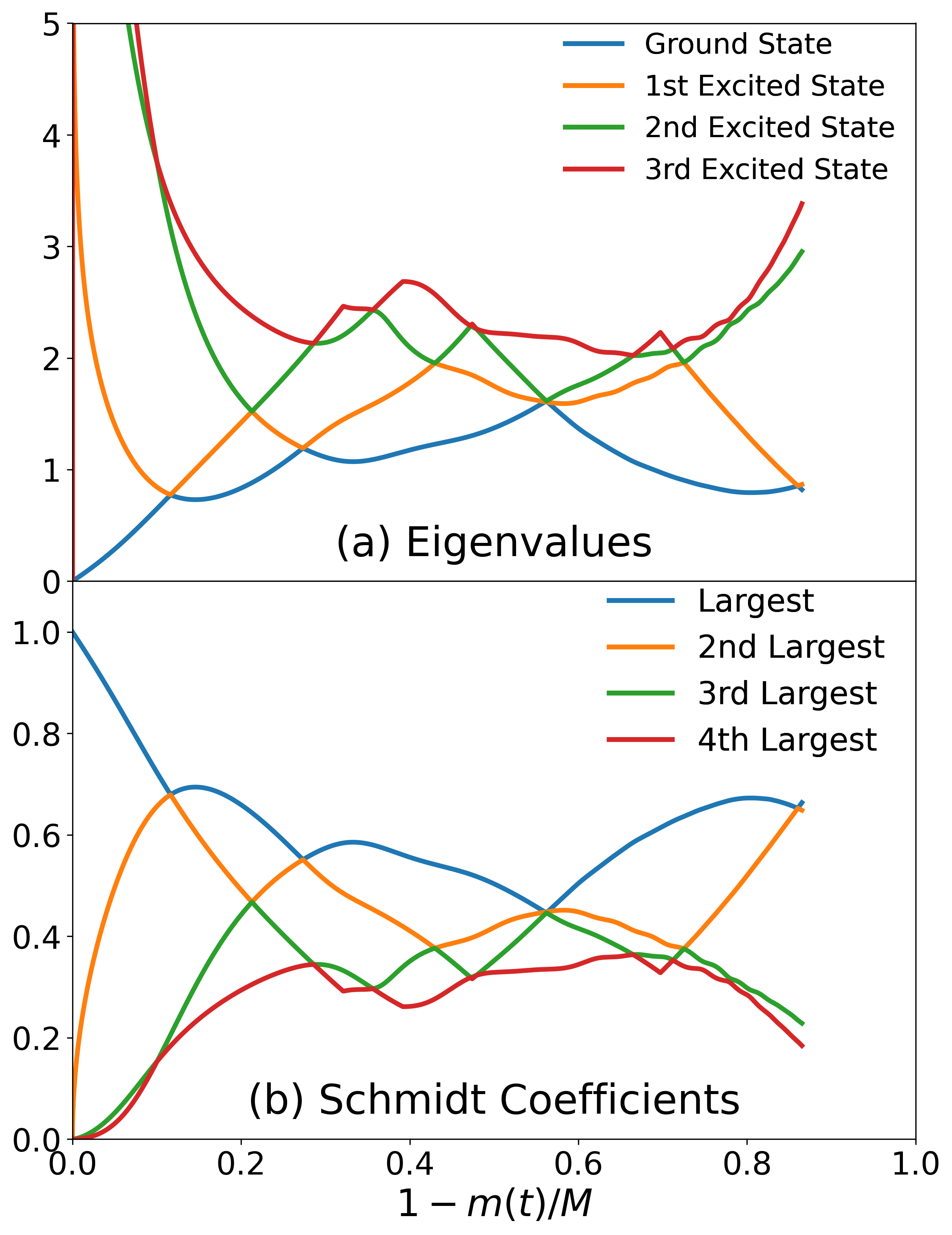}
    \caption{(a) MPS results for the time evolution of the lowest four eigenvalues of the time-dependent entanglement Hamiltonian [Eq.~\eqref{text_eq:ent_ham defined}] plotted agains the decayed fraction $1-m(t)/M$ and for the same parameters as in Fig.~\ref{fig:ee}, but fixing $M=5$ and $V=0.8$.
    The level crossing of the ground state with the first excited state happens at the same value of the decayed fraction as the non-analyticity observed in Fig.~\ref{fig:min-ee}, indicating that this behavior can be connected to a quantum phase transition in the entanglement Hamiltonian when tuning time as control parameter.    
    %We plot the crossing between the ground state and the first excited state observed in the \entham signifying a quantum phase transition that results in non-analyticity in min-entropy in Fig.~\ref{fig:min-ee}. The parameter values in Eq.~\eqref{eq:model} are: $M=5$, $V=0.8$, $t_s = t_e =1 $, $g=0.5$. Simulation time is from $t=0.0$ to a maximum time of $t_{\text{max}} = 40.0$ with time steps $\mathrm{d}t=0.02$. 
    %Only the (a) Eigenvalues of the \entham are plotted against decayed fraction of particles into the environment, where quantum phase transition happens at the same time as when min-entropy develops non-analyticity. 
    (b) Same as in (a), but for the largest four Schmidt coefficients, which are connected to the eigenvalues of the \entham via Eq.~\eqref{text_eq:eigenvalues of entham}. 
    %are plotted against the decayed fraction of particles into the environment. Eigenvalues are extracted from these Schmidt coefficients, as explained in Appendix \ref{subsec. entanglement hamiltonian} (Eq.~\eqref{eq:eigenvalues of entham}).
    }
    \label{fig:entham}
\end{figure}

\textit{Quantum phase transition in the entanglement Hamiltonian.---} As explained in Section \ref{sec. model} and motivated by the free case \cite{Kehrein2024Jun}, we analyze the spectrum of the \entham which is defined as 
\begin{equation}
    \Hh_E \equiv -\ln \rho_s
    \label{text_eq:ent_ham defined}
\end{equation}
where $\rho_s$ is the reduced density matrix of the system. 
The eigenvalues of the entanglement Hamiltonian are given by
\begin{equation}
    \epsilon_i = -\ln \lambda_i .
    \label{text_eq:eigenvalues of entham}
\end{equation}
where $\lambda_i$ are the Schmidt coefficients. We refer the readers to App.~\ref{subsec. entanglement hamiltonian} for further details.

\begin{figure}
\centering
%\captionsetup{justification=centering}
\includegraphics[width=0.49\textwidth]{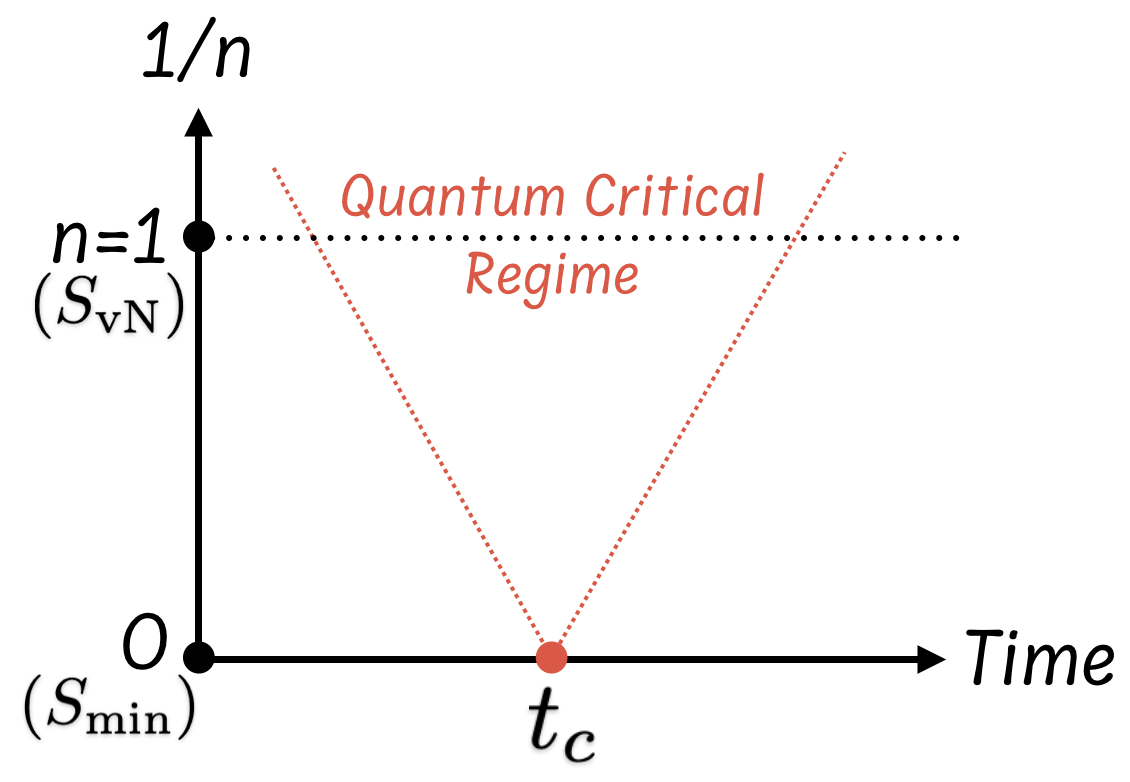}
    \caption{Schematic depicting the quantum phase transition in the \entham at a critical time $t_c$ where the min-entropy develops a non-analyticity.
    Drawing similarities with the conventional scenario for quantum phase transitions, the control parameter along the $x$-axis here is time and the role of temperature (fictitious) along the $y$-axis is here played by the inverse of the \renyi parameter $1/n$ (see Eq.~\eqref{text_eq:label_connecting_entropy_and_free_energy} as well as Appendix~\ref{subsec. entanglement hamiltonian}). 
    We associate the quantum critical point in this sketch to the non-analyticities of the temporal evolution of the min-entropy, which in this picture separates two different phases of matter during the temporal evolution of our model. Moreover, there exists a quantum critical regime at finite (fictitious) temperature, whose behavior is affected by the underlying quantum critical point at $1/n=0$. Since we find $t_c<\tpage$, the bending down of the vNEE as shown in Figs.~\ref{fig:ee_in_time} and \ref{fig:ee}, and hence the breakdown of 
    %Hawking's 
    the semiclassical picture~\cite{Calabrese2005Apr},
    %result sketched in Fig.~\ref{page}
    is associated with the presence of a critical point at an instant of time $0 \lesssim t_c \lesssim \tpage$.
    }
    \label{fig:schematic for dynamical QPT}
\end{figure}

Since the \renyi parameter $n$ serves the role of inverse fictitious temperature $\beta_{\text{fictitious }}$, we analyze the behavior of the ground state of the \entham under temporal evolution by plotting the eigenvalues of the ground state ($=$ $S_{\text{min}}$) as well as the first three excited states against the decay fraction in Fig.~\ref{fig:entham}(a). 
Since the lowest four eigenvalues are derived from the largest four Schmidt coefficients (see Eq.~\eqref{text_eq:eigenvalues of entham}), we also plot the evolution of the four largest Schmidt coefficients in Fig.~\ref{fig:entham}(b). 
In the following, we focus on the behavior of the ground state of the \entham which restricts the analysis to crossings of the ground state with the first excited state. 
We find a one-to-one correspondence between Figs.~\ref{fig:min-ee} and \ref{fig:entham} where non-analyticity in the min-entropy appears exactly when there is a crossing of the ground state with the first excited state of the entanglement Hamiltonian. 
Therefore, we provide a robust interpretation as found for the free case \cite{Kehrein2024Jun} that also holds when interaction is switched on: the system undergoes a quantum phase transition in the \entham at the same instance of time when the min-entropy develops a non-analyticity. 
As explained in Appendix~\ref{subsec. entanglement hamiltonian} (Eq.~\eqref{eq:label_connecting_entropy_and_free_energy}), the connection between the \renyi entropies $S_n$ and the thermodynamic free energy $\Ff_E(n)$ corresponding to the \entham is given by
\begin{equation}
    S_{n} = \frac{n}{n-1} \Ff_E(n) \,.
\label{text_eq:label_connecting_entropy_and_free_energy}
\end{equation}
This allows us to construct a mapping of the non-analytical behavior in time (``temporal entanglement transition'') to quantum critical behavior of the ground state of the \entham: 
we extend the usual picture of continuous quantum phase transitions to the present situation. The quantum critical point at zero temperature corresponds to the critical behavior of $S_{\rm min}$ at the instant of time $t_c$. Similar to standard quantum phase transitions, we expect a quantum critical regime at fictitious temperature; i.e., for the \renyi entropies with finite $n$, we do not expect a sharp critical point as a function of time, but a crossover between two different phases, with the behavior in the crossover region determined by the properties of the quantum critical point observed in $S_{\rm min}(t_c)$. This is illustrated in Fig.~\ref{fig:schematic for dynamical QPT}. This leads to the following natural interpretation: the bending down of the vNEE to obtain the Page curve in Figs.~\ref{fig:ee_in_time} and \ref{fig:ee} at a later time than the critical time ($\tpage>t_c$) is due to the quantum critical regime and, thus, is affected by the presence of an underlying quantum critical point. 
Therefore, we expect the generic picture to hold for arbitrary systems, that the non-analyticity in the min-entropy separates two different quantum phases of matter, which leads to a break-down of the semi-classical picture \cite{Calabrese2005Apr}. 

\begin{comment}
\begin{table}[h]
    \centering
    \caption{Table with Multirow Entries}
    \begin{tabular}{|c|c|}
        \hline
        \multirow{2}{*}{Column 1} & Column 2 \\ 
        \cline{2-2}
                                   & Text 1 \\ 
                                   & Text 2 \\ 
        \hline
        Entry 2                     & Corresponding Text A \\ 
        \hline
        Entry 3                     & Corresponding Text B \\ 
        \hline
    \end{tabular}
\end{table}
\end{comment}

%\section{Mesoscopic systems: time-dependent entanglement Hamiltonian and conserved charge picture}

\section{Time-dependent entanglement Hamiltonian and the conserved charge picture}
\label{sec. conserved charge picture}
We generalized the results of Ref.~\cite{Kehrein2024Jun} for the free case to interacting systems with conserved charge, where we found a robust behavior of the non-analyticity in the min-entropy and the corresponding temporal quantum phase transition in the entanglement Hamiltonian. 
The \entham $\Hh_E(t)$ conserves the particle number, thereby having disconnected conserved particle number sectors. Hence, any change of particle number in the system as a function of time is associated to a level crossing of the actual ground state with another eigenstate with different number of particles.
As the system empties out into the environment, so does the ground state of $\Hh_E$. 
The reduced density matrix of the system $\rho_s(t)$, which is used to define the \entham (see Eq.~\eqref{text_eq:ent_ham defined}), is pushed into smaller effective Hilbert space with time. 
Since we start with a product state \eqref{eq:initial state}, at time $t=0$ the net von Neumann bipartite entanglement entropy is zero and the unitary evolution preserves the total entropy of the system plus the environment. 
Accordingly, the von Neumann entropy of the environment remains bounded by the logarithm of the dimension of the effective Hilbert space of the system in the infinite time limit. 
The \entham is initially gapped, and the gap closes as the critical time is approached at which one particle is emitted into the environment. 
The ground state energy (equivalently the min-entropy) develops a non-analyticity at the crossing, and therefore the critical point corresponds to the quantum phase transition in $\Hh_E(t)$ described in the previous section.
Therefore, for the present system, the conserved charge picture is crucially at the center for understanding the critical behavior that develops in time. 
Our results show that the phenomenon is robust for weak to intermediate interacting strengths. 
The reason for this is that in this model, the interaction in the system does not affect the existence of the disconnected conserved charge sectors and as the particle is emitted, there is the aforementioned level crossing, e.g., between the $M$ particle sector and the $M-1$ particle sector leading to the nonanalytical behavior as a function of time. 
This conserved charge picture holds 
%Our findings and reasoning further show that 
independent of the interactions as well as the dimensionality of the model with a conserved charge; therefore, there must exist a non-analyticity in min-entropy and the quantum phase transition in $\Hh_E(t)$ at least for finite systems. 
However, the discussion so far has focused on finite systems, but the scenario sketched in Fig.~\ref{fig:schematic for dynamical QPT} only holds in the TL.
%the temporal quantum phase transition}. 
%Therefore, the quantum critical point (time in our case) divides two different quantum phases of matter where semi-classical picture breaks down and so does perturbative analysis starting from the initial time. Our findings and reasoning further show that independent of the interactions as well as the dimensionality of the model with a conserved charge, there must exist a non-analyticity in min-entropy and the quantum phase transition in $\Hh_E(t)$ at least for finite systems. As is evident from Fig.~\ref{fig:schematic for dynamical QPT}, the quantum phase transition in the \entham dynamically develops with temporal evolution. Since the \entham conserves the total particle number, the order parameter is the change in particle density (in other words, the conserved sectors where the ground state of the \entham starting from $M$ particle sector emits a particle into the environment and goes into $M-1$ particle sector). Therefore, there is a \textit{depletion transition} that drives the quantum phase transition. Since the behavior is linear up to the first non-analyticity in min-entropy, we proceed to extrapolating to the thermodynamic limit using linear regression in the next section.
Therefore, the question arises, if this level crossing scenario observed for finite systems leads to non-analytical behavior in the TL, which would manifest true quantum critical behavior. If this is the case, then the question for the order parameter arises. Here, the change in particle number discussed for the finite systems should lead to a depletion transition in the TL, at which the filling of the system serves as order parameter. As for the non-interacting system~\cite{Kehrein2024Jun}, we expect in the TL a continuous change of this order parameter with time, so that the general theory of continuous phase transitions and the overall behavior sketched in Fig.~\ref{fig:schematic for dynamical QPT} will be applicable. This scenario is further analyzed in the next section by explicitly performing a scaling analysis to the TL using our numerical MPS results for different interaction strengths.

\section{Finite Size Scaling and Critical Exponent}
\label{sec. exponent}

We start by doing a finite size scaling and then evaluating the critical exponent $\beta$ corresponding to the order parameter for the transition, namely the difference of particles in the environment. We first proceed by explaining the scaling relations and then present the results for our case. The parameters are the same as considered throughout this work: $t_s = t_e = 1.0$ as well as the system-environment coupling $g=0.5$. Starting with the critical time $t_c$ (which vanishes in the thermoydnamic limit for the system-environment coupling $g=0.5$, as shown in Section \ref{sec. regression}), we have the following scaling ansatz to illustrate that indeed there is a linear dependence on $1/M$ (thereby justifying the usage of linear regression in Section \ref{sec. regression}):
\begin{equation}
    t_c \sim M^{-1/c}
\end{equation}
where $c$ is the scaling ansatz exponent and $M$ is the number of sites in the system. Then once we have obtained the exponent $c$, we have the finite size scaling for the order parameter $(1-m(t)/M)$ which is the decayed fraction of particles to the environment
\begin{equation}
    (1-m(t)/M) M^{b/c} \sim (t-t_c)M^{1/c}
\end{equation}
where we are in the small neighborhood of $t_c$. Notice that the transition happens when the first particle leaks into the environment. That's why the order parameter is zero for $t<t_c$ while it becomes non-zero for $t>t_c$ in the thermodynamic limit (see Section \ref{sec. conserved charge picture} above for a detailed discussion). Using this scaling ansatz, we can extract the scaling ansatz exponent $b$. Finally, the ground-state energy of the \entham develops non-analyticity in time and we perform a scaling analysis in the neighborhood of the non-analytic point as
\begin{equation}
    \Big( \frac{S_\text{min}}{M} \Big) M^a \sim (t-t_c)M^{1/c}
\end{equation}
which allows us to calculate the scaling ansatz exponent $a$. Once we have performed the scaling, we expect a perfect collapse of the order parameter as well as the min-entropy for various system sizes $M = \{3,4,5,6,7\}$. Indeed, we find universal exponents $a, b$ and $c$ such that \textit{for all values} of $V = \{0.0, 0.4, 0.8, 1.2\}$, we find a perfect collapse. The free case $V=0.0$ is used to benchmark against the analytical calculations performed in the thermodynamic limit in Ref.~\cite{Kehrein2024Jun}. We provide one such example of collapse here for the case of $V=1.2$ in Fig.~\ref{fig:collapse plot V=1.2} while other plots for different values of $V$ are provided in Appendix \ref{app. exponent}. We find the values of all three scaling ansatz exponents ($a, b, c$) to be around $1.0$. The precise values are shown in Table \ref{table:exponents}. This shows a consistent collapse for the order parameter as well as the non-analyticity in the ground state energy of the entanglement Hamiltonian.

Then we find the \textit{critical exponent} $\beta$ of the order parameter in the small neighborhood of $t_c$ such that $t>t_c$ which is given by
\begin{equation}
    \Big( 1 - \frac{m(t)}{M} \Big) = A(t-t_c)^{\beta} \qquad (t>t_c).
    \label{eq:critical exponent lambda}
\end{equation}
For the free case, in the thermodynamic limit, the critical exponent is found to be $\beta = 0.5$ as shown in Eq.~(61) of Ref. \cite{Kehrein2024Jun}. For finite systems, we fix the value of $M$ and then perform the scaling to extract the values of $A$ and $\beta$. We know that the prefactor $A$ is not universal, so we restrict our analysis for $\beta$. Then for a given interaction strength $V$ and fixed total number of sites $L$, we aggregate the value of $\beta$ for $M=\{3,4,5,6,7\}$. Within reasonable accuracy, we find the universal critical exponent $\beta$ as the free case \cite{Kehrein2024Jun} for all values of $V$. We provide one such example for $M=5$ and $V=0.8$ in Fig.~\ref{fig:critical exponent M=5, V=0.8}. Aggregated values for various $V$ are summarized in Table~\ref{table:lambda exponent}. Therefore we conclude \textit{the robustness of such temporal quantum phase transitions} that the interactions don't change the critical exponent associated with the order parameter .

\onecolumngrid

\begin{figure}
\centering
%\captionsetup{justification=centering}
\includegraphics[width=\textwidth]{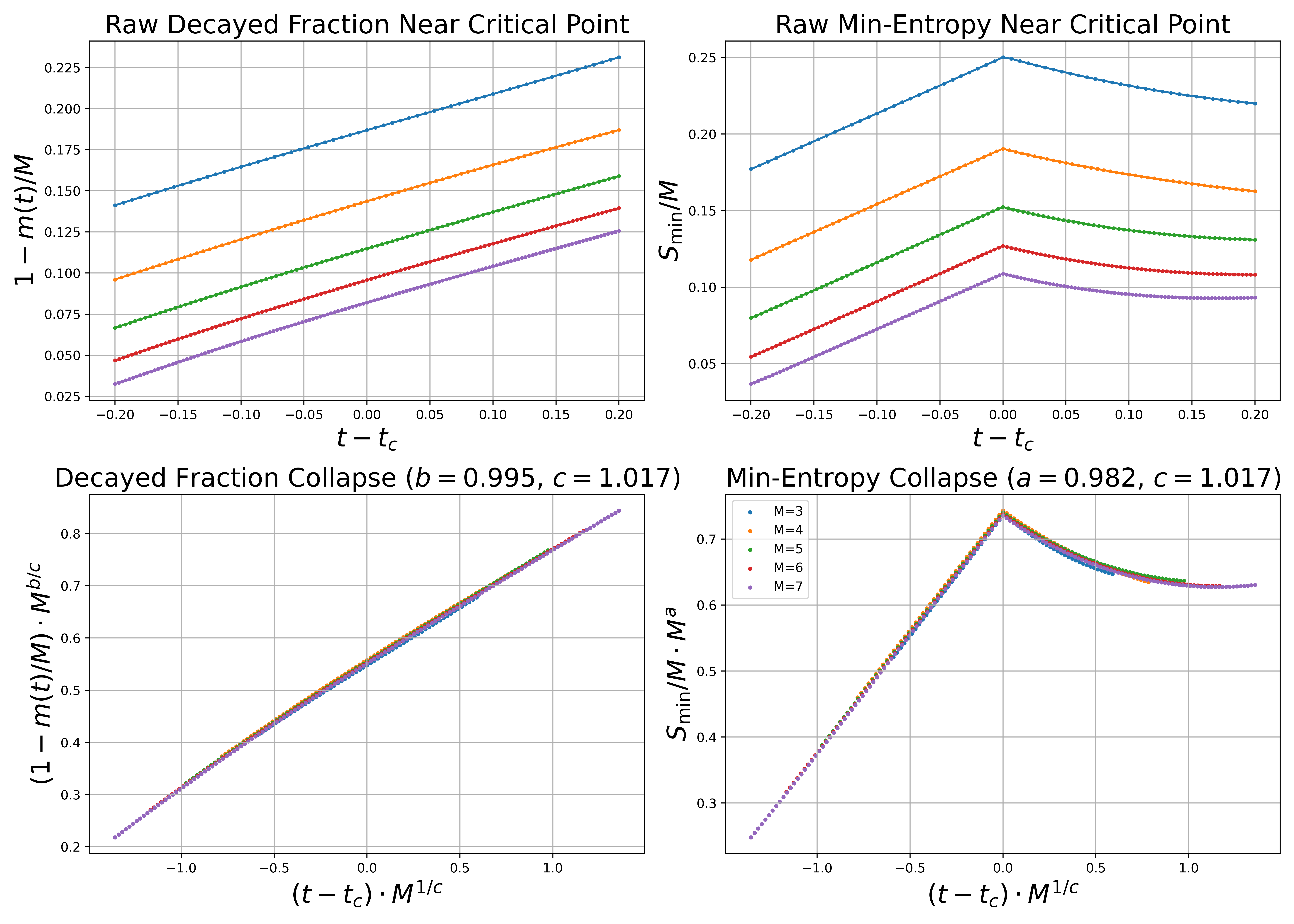}
    \caption{Top row: raw data in the vicinity of the critical point where top left is for the decayed fraction of particles against time, while the top right plots min-entropy against time (axes are re-scaled system size as shown in the plots). Bottom row: finite-size scaling for the raw plots in the neighborhood of the critical (non-analytic) point, showing a perfect collapse. Using the scaling relations explained in Section \ref{sec. exponent}, we find universal values of scaling ansatz exponents $a, b$ and $c$ \textit{for all values of} $V$. The plot is shown for $V=1.2$, $g=0.5$, $t_s = t_e = 1.0$, $L=50$ and $M=\{3,4,5,6,7\}$. Plots for other values of $V$ are shown in Appendix \ref{app. exponent}. Therefore, we have a consistent collapse for both the order parameter as well as the ground state energy of the entanglement Hamiltonian.}
    \label{fig:collapse plot V=1.2}
\end{figure}

\twocolumngrid

\begin{table}
    \centering
    \caption{Scaling exponents, as explained in Section \ref{sec. exponent}, for various values of $V$. Other parameter values are $g=0.5$, $t_s = t_e = 1.0$, $L=50$ and $M=\{3,4,5,6,7\}$.}
    \label{table:exponents}
    \begin{tabular}{|c|c|c|c|c|c|c|c|}
        \hline
        V & $c$ & $b$ & $a$ \\
        \hline
        $0.0$ & $1.01$ & $0.99$ & $0.99$\\ 
        \hline
        $0.4$ & $1.02$ & $0.99$ & $0.98$\\ 
        \hline
        $0.8$ & $1.02$ & $0.98$ & $0.97$\\ 
        \hline
        $1.2$ & $1.02$ & $0.99$ & $0.98$\\ 
        \hline
    \end{tabular}
\end{table}

\begin{figure}
\centering
%\captionsetup{justification=centering}
\includegraphics[width=0.49\textwidth]{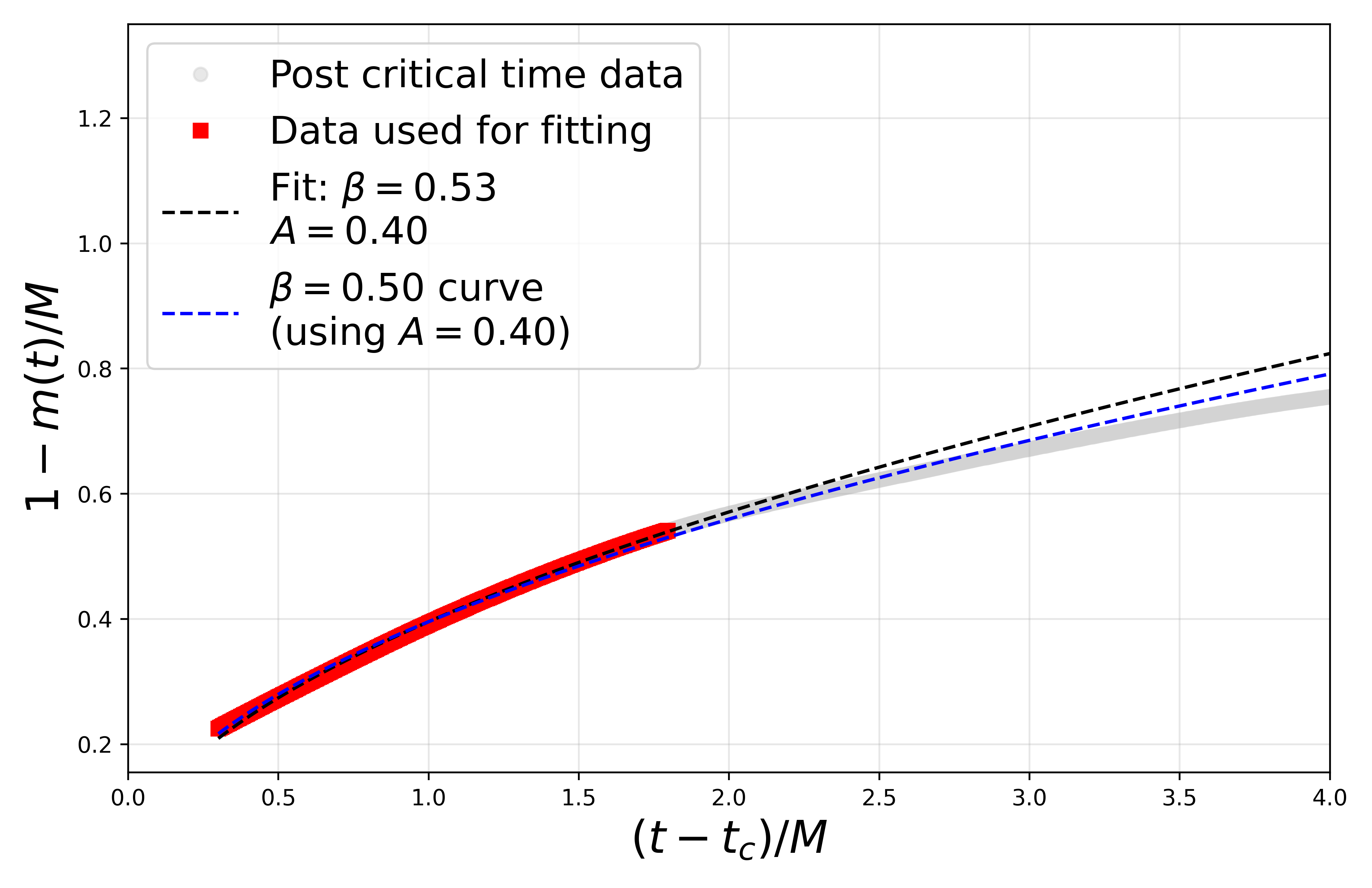}
    \caption{We perform the scaling relation in Eq.~\eqref{eq:critical exponent lambda} for $M=4$ and $V=0.8$ to get extract the critical exponent $\beta$ associated to the order parameter. The post-critical time data is used that is in a small neighborhood of the critical time. The post-critical time data is shown for clarity purposes where we expect the scaling relations to fail as they are valid only in the vicinity of the critical time. The black dashed line denotes the actual fit while the blue dashed line denotes the curve in Eq. \eqref{eq:critical exponent lambda} for $\beta = 0.50$ while using the same value of $A$ as obtained via fitting in the black dashed curve. Accordingly, we find to a reasonable accuracy the value of the critical exponent $\beta$ same as that of the free case in the thermodynamic limit as shown in Eq.~(61) in Ref.~\cite{Kehrein2024Jun}. Other parameters are: $L=50$, $g=0.5$ and $t_s = t_e = 1.0$. We perform an average over different $M=\{3,4,5,6,7\}$ for a fixed $V$ to get aggregated value for finite systems and the results are summarized in Table~\ref{table:lambda exponent}.}
    \label{fig:critical exponent M=5, V=0.8}
\end{figure}

\begin{table}
    \centering
    \caption{Critical exponent, as explained in Eq.~\eqref{eq:critical exponent lambda}, for various values of $V$. Other parameter values are $g=0.5$, $t_s = t_e = 1.0$ and $L=50$. We have averaged for $M=\{3,4,5,6,7\}$ to get a value for finite systems and within reasonable accuracy, we find $\beta \approx 1/2$ --- the same as the free case $V=0.0$ in the thermodynamic limit \cite{Kehrein2024Jun}. Accordingly, interactions don't affect the robustness of such temporal quantum phase transitions.}
    \label{table:lambda exponent}
    \begin{tabular}{|c|c|c|c|c|c|c|c|}
        \hline
        V & $\beta$ & $A$  \\
        \hline
        $0.0$ & $0.51$ & $0.42$ \\ 
        \hline
        $0.4$ & $0.52$ & $0.40$ \\ 
        \hline
        $0.8$ & $0.54$ & $0.37$ \\ 
        \hline
        $1.2$ & $0.52$ & $0.33$ \\ 
        \hline
    \end{tabular}
\end{table}

 \section{Analysis in the thermodynamic limit}
  \label{sec. regression}

\begin{figure}
\centering
%\captionsetup{justification=centering}
\includegraphics[width=0.49\textwidth]{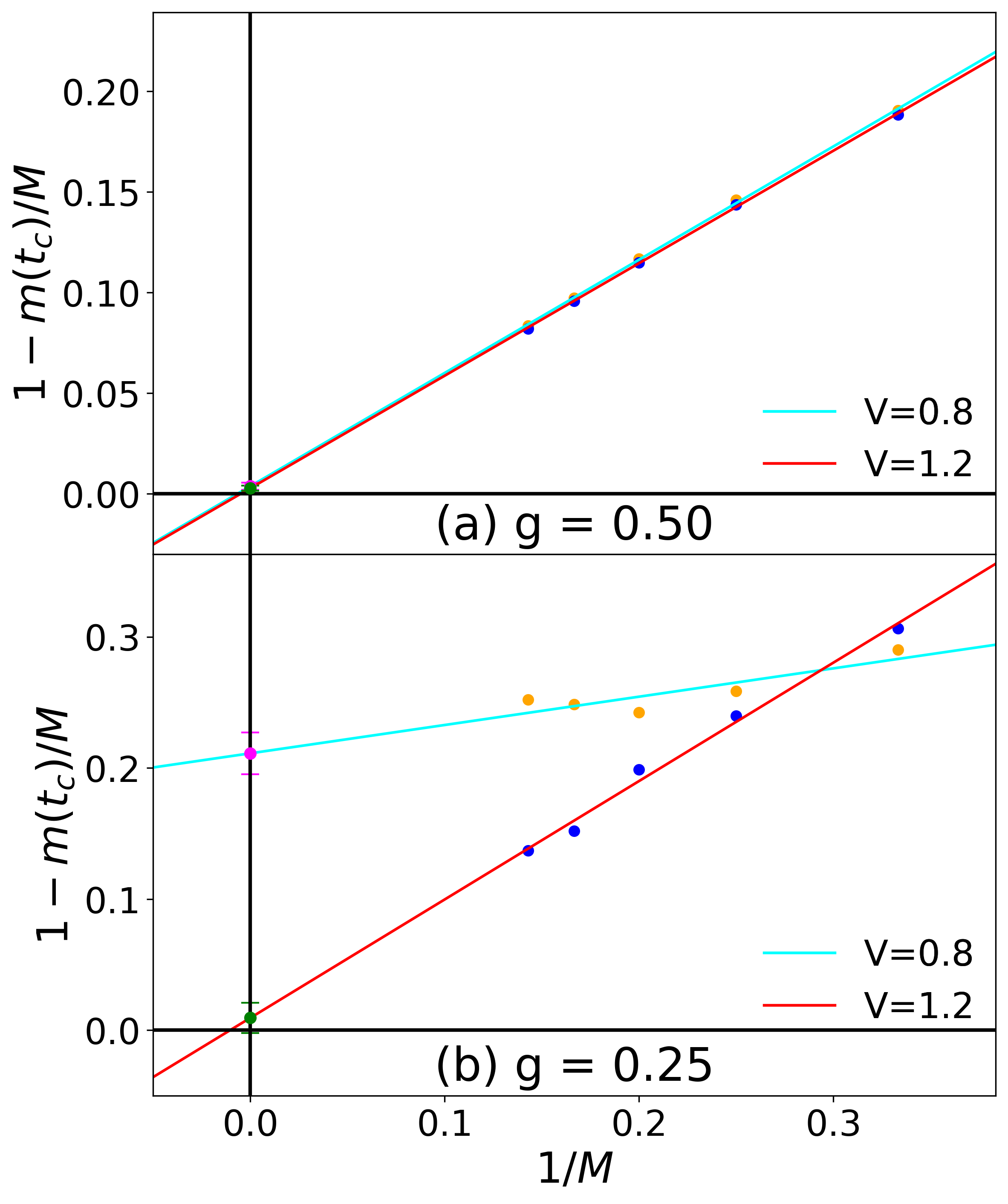}
    \caption{Linear regression plot where we extrapolate the critical decayed fraction corresponding to the quantum critical time (as shown in Fig.~\ref{fig:schematic for dynamical QPT}) to the thermodynamic limit where $M \to \infty$. The system sizes are varied from $M=3$ to $M=7$ for two interaction strengths (total of $5$ data points each) as shown in the bottom plots of Fig.~\ref{fig:min-ee} for $V=0.8$ and Fig.~\ref{fig:appendix_min_ent} for $V=1.2$. Other parameters are $t_s=t_e=1.0$ where the coupling $g$ between the system and the environment is shown in the plot. We find that the critical decayed fraction vanishes in the thermodynamic limit for the stronger coupling $g=0.5$ between the system and the environment, however for weaker coupling $g=0.25$, there exists a non-zero value in the thermodynamic limit. As discussed in the main text, we find an acute sensitivity to the coupling strength which is summarized in Table \ref{table:regression_critical_decay_fraction} where even for the weaker coupling, the thermodynamic critical value vanishes with increasing interaction strength $V$.}
    \label{fig:regression plot}
\end{figure}

We discuss a scaling analysis by performing a linear extrapolation to the thermodynamic limit (as justified above in Section \ref{sec. exponent}). 
Considering that the short time behavior of the min-entropy until the first non-analyticity is effectively linear, we expect this to capture the leading order of the behavior. As motivated by the free case \cite{Kehrein2024Jun}, we expect later non-analyticities to smear out in the thermodynamic limit. For a fixed interaction strength $V=0.8$ (as shown in Fig.~\ref{fig:min-ee}), we select the $x$-coordinates of the first non-analytic points for different system sizes that we denote as $1-m(t_c)/M$ and perform a linear regression as shown in Fig.~\ref{fig:regression plot}. The system parameters treated and the results are summarized in Table \ref{table:regression_critical_decay_fraction} and a CSV file containing a summary of regression analyses is provided in the research data management \cite{zenodo}.
We find an \textit{acute sensitivity on the coupling strength} $g$ between the system and the environment, where we find that the non-vanishing critical value persists for higher interaction strengths when the coupling is weaker. 
For both cases of the coupling ($g=0.5$ and $g=0.25$), the value corresponding to the free case $V=0.0$ indeed matches the value obtained in the exactly solvable free model as carried out in Ref.~\cite{Kehrein2024Jun}. At finite $V>0$, for the smaller value of $g$ we find a non-vanishing $t_c$-value in the TL for $V \lesssim 1$. 
However, when further increasing $V$, or in the case of $g=0.5$, we always obtain $t_c \approx 0$; a conservative estimate of the numerical errors after the extrapolation gives an errorbar of $\sim 3$ times the standard deviation shown, so that we conclude that these values are zero within our obtained numerical accuracy. This is also true at $V=0$, in agreement with Ref.~\cite{Kehrein2024Jun} (see Table \ref{table:regression_critical_decay_fraction}). 
Note that we have performed regression analyses for the \textit{critical value of decay fraction} instead of \textit{critical time} (as plotted in Fig.~\ref{fig:ee_in_time} for the vNEE); the scaling analyses for the critical time is provided in the Appendix~\ref{appendix} where we observe the same pattern.

\begin{table}
    \centering
    \caption{Linear regression for the first non-analytic points across different system sizes in Fig.~\ref{fig:min-ee} that corresponds to the quantum phase transition in the \entham (crossing shown in Fig.~\ref{fig:entham}). In the regression, the $Y-$axis is $1-m(t_c)/M$ where $m(t)$ is the number of particles in the system at given time $t$ and the critical time is denoted by $t_c$ where non-analyticity develops in min-entropy. The $X-$axis is $1/M$, therefore the $Y-$intercept is the extrapolation of the critical decayed fraction in the thermodynamic limit. For intermediate to stronger interactions, the thermodynamic value for decayed fraction vanishes and we provide an analysis in the text based on Fig.~\ref{brick-wall}. We also note an acute sensitivity on the tunneling strength $g$ where the thermodynamic value vanishes at intermediate interactions for higher value of $g$. A similar analysis but for critical time instead of critical decayed fraction is performed in the Appendix \ref{appendix}.}
    \label{table:regression_critical_decay_fraction}
    \begin{tabular}{|c|c|c|c|c|c|c|c|}
        \hline
        V & g & $Y-$intercept & Slope & $R^2$ \\
        \hline
        \multirow{2}{*}{$0.0$} & $0.50$ & $0.0001 \pm 0.0001 $ & $0.5975 \pm 0.0003 $& $1.0000 $\\  
         & $0.25$ & $0.2035 \pm 0.0283 $& $0.3481 \pm 0.1237 $& $0.7252 $\\ 
        \hline
        \multirow{2}{*}{$0.4$} & $0.50$ & $0.0020 \pm 0.0009 $ & $0.5847 \pm 0.0041 $& $0.9999  $\\
        & $0.25$ & $0.2091 \pm 0.0092 $& $0.2696 \pm 0.0402 $& $0.9374  $\\ 
        \hline
        \multirow{2}{*}{$0.8$} & $0.50$ & $0.0037 \pm 0.0018  $ & $0.5628 \pm 0.0078  $& $0.9994   $\\
        & $0.25$ & $0.2111 \pm 0.0159 $& $0.2159 \pm 0.0695 $& $0.7627   $\\ 
        \hline
        \multirow{2}{*}{$1.2$} & $0.50$ & $0.0027 \pm 0.0013  $ & $0.5590 \pm  0.0058 $& $0.9997   $\\
        & $0.25$ & $0.0094 \pm 0.0116 $& $0.9030 \pm 0.0509 $& $0.9906  $\\ 
        \hline
    \end{tabular}
\end{table}

\begin{figure}
\centering
%\captionsetup{justification=centering}
\includegraphics[width=0.49\textwidth]{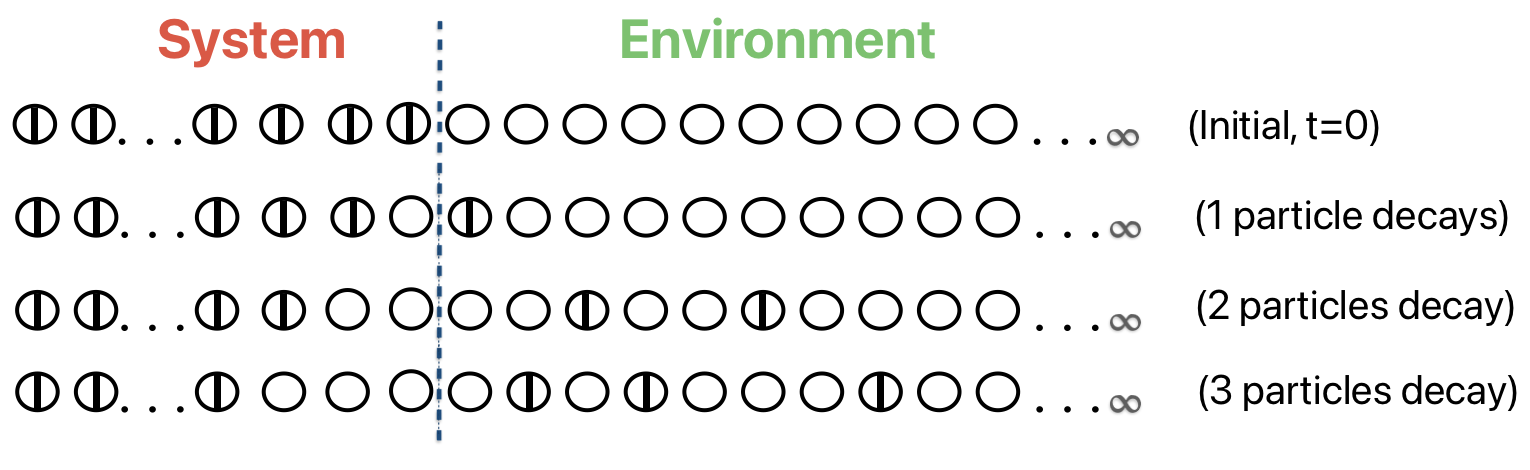}
    \caption{When a particle leaks to the environment, a hole enters. Energetically favorable process is where holes form a domain-wall structure via higher order processes. In the case of strong interacting limit where $V/t_h\gg 1$, only one particle leaks and the system gets stuck. This allows to write the ansatz as in Eq.~\eqref{eq:ansatz}. We refer to the main text for more details.}
    \label{brick-wall}
\end{figure}

\textit{Analysis.---} We now provide an interpretation and analysis for the observed vanishing of the thermodynamic critical value of decay fraction as the interaction strength is tuned up. At $t=0$, we start with the initial state \eqref{eq:initial state}. Then a particle leaks to the environment, or equivalently, a hole enters the system. Now the particle is free to hop in the environment but the hole cannot penetrate deeper into the system due to energy constraints imposed by the interaction term in model \eqref{eq:model}. 
For a hole to move one site deeper into the system, there is an energy cost of $V$. However, there is another energetically favorable scenario where higher order processes play a role to emit another particle into the environment (a second hole coming into the system) without the first hole hopping any further into the system. This leads to a domain-wall structure of holes entering the system. This is explained schematically in Fig.~\ref{brick-wall}.

We can make this interpretation concrete to show the vanishing of the critical values in the thermodynamic limit. Consider the strongly interacting case where $V/t_h \gg 1$ (where $t_h \equiv t_s = t_e$ which has been set to $1.0$ in this work). 
Contrary to intuition, repulsive interactions don't generically lead to particles spreading out. The concept of pairing induced by repulsive interactions is known in the literature \cite{repulsive-bound-pair-1, Heidrich-Meisner2009Oct, repulsive-bound-pair-2, repulsive-bound-pair-3, repulsive-bound-pair-4, repulsive-bound-pair-5} where repulsively bound pairs form as a consequence of repulsive interactions . In fact this behavior is prevalent in our setup too where a strong value of $V$, say $V=5.0$ given other parameters are same as in our paper ($g=0.5$, $t_s=t_e = 1.0$, $L=50$ and $M$ can be anywhere between $3$ and $7$ sites) causes the system to freeze and no particles leak into the environment. See Appendix \ref{app. strongly interacting case} for $V=3.0$ and $V=5.0$ for time scales accessible to us. Accordingly, in order to write down an ansatz for the number of particles in the system, we have two possibilities: (1) one particle decays to the environment and accordingly a hole enters the system, causing dynamics in the setup \footnote{In this strongly interacting case, no higher order processes will play any relevant role and the system will freeze in this state and no further decay of particles will be observed.}, and (2) no particle leaks into the environment, thereby having no dynamics at all (see Appendix \ref{app. strongly interacting case}).
Since we are interested in finding the crossing of the ground state of the entanglement Hamiltonian and the dynamics is dominated by these two possibilities, only the largest and the second-largest Schmidt coefficients, namely $1-\lambda(t)$ and $\lambda(t)$ respectively, are playing a role such that $\lambda(t=0) = 0$ (see Appendix~\ref{app. strongly interacting case} where we have also provided numerical evidences for Schmidt values). The ansatz reads as
\begin{equation}
     m(t) = \frac{M  e^{2\ln(1-\lambda(t))} + (M-1)  e^{2\ln(\lambda(t))}}{e^{2\ln(1-\lambda(t))} + e^{2\ln(\lambda(t))}}
\label{eq:ansatz}
\end{equation}
where we analyze a parametric plot of the Schmidt coefficients $\lambda(t)$ and $1- \lambda(t)$ on the y-axis and the decay fraction $1-(m(t)/M)$ on the x-axis. By increasing the system size $M$ to as large as possible ($\sim 10^6$), we find the critical value of decay fraction when the two Schmidt coefficients cross to be vanishingly small ($\sim 10^{-7}$). A Mathematica file providing the parametric plot is provided in the research data management \cite{zenodo}. This shows that the critical time in Fig.~\ref{fig:schematic for dynamical QPT} is vanishingly small in the thermodynamic limit for strongly interacting system ($V/t_h \gg 1$) which is consistent with our regression analyses in Table \ref{table:regression_critical_decay_fraction}.

\begin{comment}
\begin{figure}
\centering
%\captionsetup{justification=centering}
\includegraphics[width=0.45\textwidth]{plots/EE_mEE_joined_joined_g0.25_0.5.png}
    \caption{We illustrate the sensitivity on the tunneling strength $g$ in Eq.~\eqref{eq:model}. Parameters chosen are $M=4$, $V=0.8$, $t_s=t_e =1.0$ where simulation is run from time $t=0.0$ to $t=40.0$. (a) Von Neumann entanglement entropy and (b) min-entropy are compared for two values of $g$ where we observe that lowering $g$ leads to more entanglement between the system and the environment. Finally as shown in Table \rj{table}, the thermodynamic extrapolation of the critical time persists to be non-zero for stronger values of interaction strength for lower value of $g$.}
    \label{tunneling-sensitivity}
\end{figure}
\end{comment}

 \section{Summary and Outlook}
 \label{sec. conclusion}

We studied the temporal evolution of entanglement dynamics between bipartite subsystems (the system and the environment, see Fig.~\ref{setup}) of an interacting spinless fermionic chain. 
Motivated by the construction in black hole physics, we made the environment much larger than the system and initiated the time evolution from a product state (Eq.~\eqref{eq:initial state}) where the system was completely filled while the environment was empty. 
While in Ref.~\cite{Kehrein2024Jun} the non-interacting system was considered, we here extend the picture to the interacting case.
This is done by adding interactions in the system, while the environment remained non-interacting. 
%The free case has already been studied in Ref. whose results are generalized in this work. 

Contrary to a typical many-body system where the vNEE displays a linear behavior followed by a volume law saturation \cite{Calabrese2005Apr}, we found a Page curve dynamics in the vNEE (Figs.~\ref{fig:ee_in_time} and \ref{fig:ee}) where there is a bending down at the Page time. We then analyzed the spectrum of the \entham associated with the system 
%(details are in Appendix \ref{subsec. entanglement hamiltonian}) 
whose ground state energy is given by min-entropy (Eq.~\eqref{min-entropy definition}). Therefore, we studied the dynamical evolution of min-entropy (Fig.~\ref{fig:min-ee}) where we found a non-analyticity that developed in time. To find the cause for this non-analyticity, we analyzed the spectrum of the \entham associated with a fictitious temperature $1/n$ 
%(see Appendix \ref{subsec. entanglement hamiltonian}) 
where $n$ is the \renyi parameter where we found a level crossing between the ground state and the first excited state of the entanglement Hamiltonian (Fig.~\ref{fig:entham}). 
The non-analyticity developed in the ground state energy corresponds to the quantum phase transition in the \entham at critical time $t_c$. 
This led us to the picture depicted in Fig.~\ref{fig:schematic for dynamical QPT} of a temporal quantum phase transition and an associated quantum critical regime.
The quantum critical time $t_c$ divides two different quantum phases of matter that correspond to different conserved charge sectors, and any semi-classical picture or perturbative analysis starting from the initial time breaks down near the critical time at vanishing fictitious temperature.
The vNEE is analytic; however, the bending down at the Page time ($ \tpage > t_c$) might be influenced by the quantum critical regime that corresponds to an underlying quantum critical point. 
This is further illustrated by our findings in Sec.~\ref{sec. conserved charge picture} where we explained that any model with a conserved charge, independent of interaction as well as dimensionality, will exhibit such a critical behavior at least for a finite system size. 
%The conserved charge picture is at the heart of the dynamical picture obtained in this work.

\begin{figure}
\centering
%\captionsetup{justification=centering}
\includegraphics[width=0.45\textwidth]{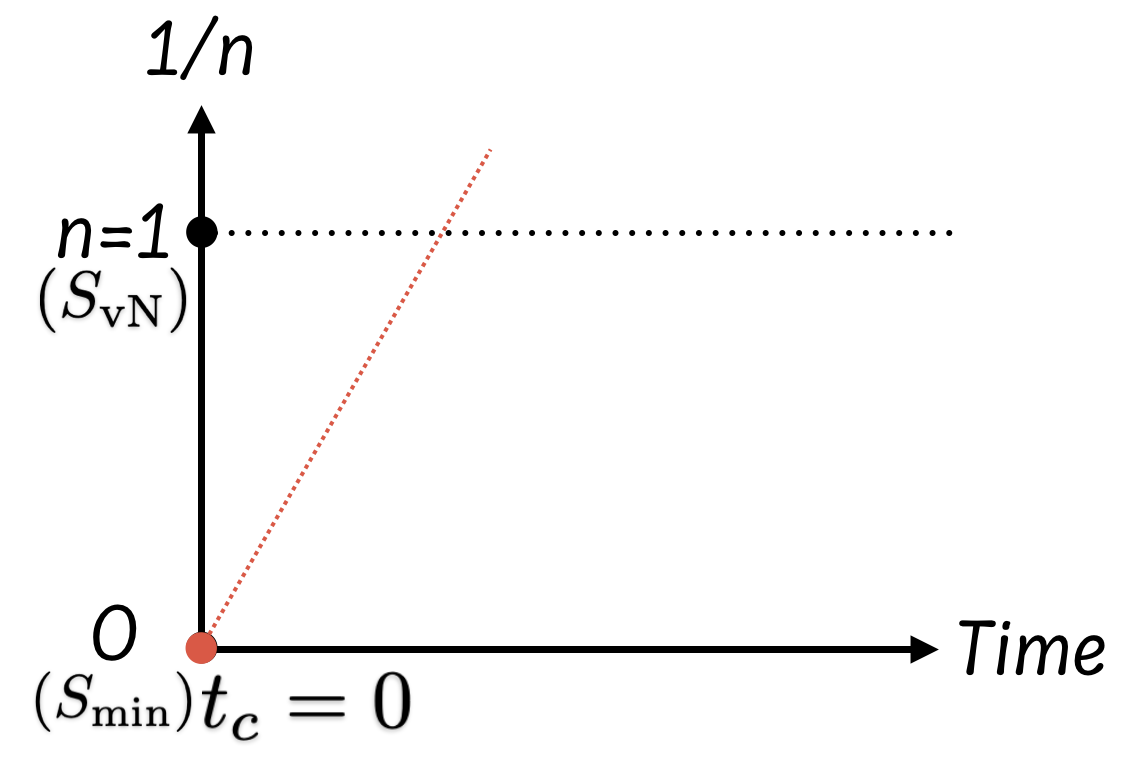}
    \caption{A schematic showing \textit{instantaneous} dynamics in the thermodynamic limit leading to the quantum phase transition in the \entham at a critical time $t_c=0$ where the min-entropy develops non-analyticity. Even though the vNEE is analytic, performing a perturbation theory still needs some care as the semi-classical description breaks down right from the initial time \cite{Calabrese2005Apr}.}
    \label{fig:schematic for instantaneous dynamical QPT}
\end{figure}

Along the same lines as in the free case \cite{Kehrein2024Jun}, we expect the later non-analyticites (after the first one) to smear out in the TL. 
Using the linear growth of the min-entropy up to the first non-analyticity, we used linear regression to extrapolate to the thermodynamic limit where we found that there exists a non-zero critical time for weak interactions at which the quantum phase transition in the \entham develops.
However, as the interaction strength increases, the critical value vanishes in the thermodynamic limit, implying an \textit{instantaneous} dynamics leading up to the non-analytic behavior (see Fig \ref{fig:schematic for instantaneous dynamical QPT}). 
Note that also for $t_c=0$ the semiclassical description loses its applicability, so that, as long as $t_c \lesssim \tpage$, the proposed scenario keeps its validity (see Fig. \ref{fig:schematic for instantaneous dynamical QPT}).
Our analyses show that one must be careful before naively applying perturbation theory by switching on the interactions because of the underlying quantum critical point that might influence the behavior of the system even for finite \renyi entropies. 
Moreover, the order parameter for the transition, whose schematic is sketched in Fig.~\ref{fig:schematic for dynamical QPT}, is the change in particle number density where a depletion transition drives the quantum phase transition. Near the critical point, the particle density develops a power law behavior with an exponent $1/2$ for the free case \cite{Kehrein2024Jun}. Since the weak interactions are nearly the same as the free case, including the thermodynamic critical value (see Table \ref{table:regression_critical_decay_fraction}), we strongly expect the power law behavior to stay the same. Indeed, as shown in details in Section \ref{sec. exponent}, we do find the same value of the critical exponent thereby concluding that such temporal quantum phase transitions are robust despite the presence of interactions which makes our setup non-integrable.

Our work is based on a weak coupling $g=0.50$ between the system in the environment but we found an interesting behavior for further weaker coupling $g=0.25$. Namely, we found an acute sensitivity on the coupling strength $g$ between the system and the environment where the weaker the coupling $g$, the stronger the interaction is required for the thermodynamic critical value to vanish. Furthermore, the critical exponent's analysis and the finite-size scaling performed in Section \ref{sec. exponent} remain \textit{no longer valid} for $g=0.25$. We do not find a universal collapse and therefore, no associated exponents. This might suggest that the nature of phase transition has changed from a continuous phase transition to a first-order transition at some critical value $g^\star$, or something completely different altogether. A true understanding for significantly smaller values of $g$ is lacking, in the sense that why it shows a finite, non-zero value of $t_c$ in the thermodynamic limit and why the exponents are no longer universal, whether it be a first-order phase transition or something completely different. This is left as an outlook.

%We established the resilience of the quantum phase transition in the \entham for fermionic chain despite switching on the interaction and argued that any model with conserved charge picture must develop a quantum critical behavior independent of dimensionality. 
%A natural question to ask is how robust such behaviors are in models without any conserved charges. If the non-analyticity in min-entropy persists, unlike the conserved charge picture here, the physical mechanism governing the dynamics needs to be identified. 
%Moreover, we found an acute sensitivity on the coupling strength $g$ between the system and the environment (see Tables \ref{table:regression_critical_decay_fraction} and \ref{table:appendix_regression_critical_decay_fraction}).
%It remains unclear, quantitatively speaking, why the coupling strength $g$ influences the physical dynamics significantly. 
A natural question to ask is whether non-analytic behavior can also occur in models without any conserved charges. Interestingly, results in Ref. \cite{sarang} show that this is indeed the case, but only in the thermodynamic limit. Another interesting question relates to the role of the coupling strength $g$ and why it can have such a significant impact on the dynamics.

Returning to our original motivation from the black hole information paradox, we have drawn analogies between the black hole setup and our lattice setup. Despite the obvious differences, there does exist some similarities between the two. 
A recent proposal in black hole physics has gathered attention, namely the \textit{island formula} \cite{almheiri-island-1, almheiri-island-2, almheiri-island-3}, where a quantum extremal surface develops during the temporal evolution of vNEE right before the Page time. 
The presence of islands is crucial after the Page time for the reproducibility of the Page curve. 
However, the emergence of islands is not smooth, thereby leading to a non-analytic behavior in the evolution of vNEE. 
This suggests the possibility of whether Page curves in the black hole information paradox can be understood via the mechanism highlighted in this work. 
Ours is a $1+1$ dimensional setup while black holes are $3+1$ dimensional objects. In one spatial dimension, we find non-analyticity in the ground state energy of the entanglement Hamiltonian at zero fictitious temperature associated with the entanglement Hamiltonian. The possibility we are talking about is that when one goes to higher dimensions, the criticality might not just be limited to zero fictitious temperature but expand to higher, non-zero fictitious temperatures, causing non-analyticity in not just the ground state but excited states of the entanglement Hamiltonian. This includes the von Neumann entropy that might have a non-analyticity in higher dimensions when approached via our methodology as sketched in this work. Numerics become practically impossible in higher dimensions, so the hope rests on analytical frameworks revolving around the entanglement Hamiltonian associated with a Schwarzchild black hole in AdS spacetime (prepared in a pure state as is normally the setup found in the literature, see Ref.~\cite{Almheiri2021Jul}) coupled to an empty environment. 
The question remains about the validity of our claims to black holes, and to which extent we can further understand the emergence of these islands as being related to some form of quantum phase transition in the associated entanglement (equivalently, modular) Hamiltonian.

\section*{Data and Code Availability}

All data and simulation codes are available on Zenodo
on reasonable request \cite{zenodo}.

\section*{ACKNOWLEDGMENTS}
All authors acknowledge financial support via the Deutsche Forschungsgemeinschaft (DFG, German Research Foundation) Grant No. 217133147 via SFB 1073 (Project No. B03), 
%They are also grateful for support from Deutsche Forschungsgemeinschaft (DFG, German Research Foundation) 
as well as Grants No. 436382789, and No. 493420525, via large equipment grants (GOEGrid). 
%Additionally, R.J. would like to 
R.J. thanks Karun Gadge, Heiko Georg Menzler and Vibhu Mishra for fruitful discussions.
S. K. acknowledges the Aspen Center for Physics, which is supported by National Science Foundation grant PHY-2210452, and the Kavli Institute for Theoretical Physics (KITP), supported by grants NSF PHY-1748958 and PHY-2309135, where part of this work was performed.

\bibliography{page_curve_interacting.bib}

\appendix

\section{Entanglement Hamiltonian}
\label{subsec. entanglement hamiltonian}

For any bipartite system $AB$ described by the density matrix $\rho_{AB} = |\psi_{AB} \rangle \langle \psi_{AB}|$ where $|\psi_{AB} \rangle $ is the eigenstate of the entire system, we can calculate the reduced density matrix as $\rho_{A/B} = \operatorname{Tr}_{B/A}(\rho_{AB})$. We can decompose $|\psi_{AB} \rangle $ via Schmidt decomposition as 
\begin{equation}
    |\psi_{AB} \rangle = \sum\limits_i^r \sqrt{\lambda_i} |i\rangle_A |i\rangle_B
\end{equation}
where $\lambda_i$ are the \textit{Schmidt coefficients} (also known as the \textit{Schmidt values}) and $r$ is the \textit{Schmidt rank}. \renyi entropy $S_n$ for any \renyi order $n$ is completely decided by the knowledge of Schmidt coefficients for the bipartite state as
\begin{equation}
S_n\left(\rho_A\right)=\frac{1}{1-n} \ln \left(\sum_{i=1}^r \lambda_i^n\right).
\end{equation}

An associated concept to bipartite system is that of the entanglement (or modular) Hamiltonian $\Hh_E$ which is defined as
\begin{equation}
    \Hh_E \equiv -\ln \rho_A
    \label{eq:ent_ham defined}
\end{equation}
whose eigenvalues are given by
\begin{equation}
    \epsilon_i = -\ln \lambda_i .
    \label{eq:eigenvalues of entham}
\end{equation}
Therefore, the Schmidt coefficients completely describe the spectrum of the entanglement Hamiltonian where the ground state eigenvalue is given by the largest Schmidt coefficient that turns out to be the \ment: 
\be 
\epsilon_{\text{GS}} = -\ln \lambda_{\text{max}} = S_{n \to \infty} = S_{\text{min}}.
\ee
Similarly, the first excited state is given by the second-largest Schmidt value, and so on and so forth. Accordingly, there is a one-to-one mapping of the bipartite system to that of entanglement Hamiltonian where the role of ``fictitious temperature'' for the \entham is provided by $1/n$, $n$ being the \renyi order. We can see this explicitly by writing the expression for the thermodynamic free energy corresponding to the \entham as
\begin{equation}
    \Ff_E(\beta_{\text{fictitious}}) = -\frac{1}{\beta_{\text{fictitious}}} \ln \Zz_E(\beta_{\text{fictitious}})
\end{equation}
where $\Zz_E(\beta_{\text{fictitious}})$ is the partition function of the \entham $\Hh_E$ given by
\begin{equation}
    \Zz_E(\beta_{\text{fictitious}}) = \Tr_{\Hh_{\text{sys}}} e^{-\beta_{\text{fictitious}} \Hh_E}.
\end{equation}
Here $\beta_{\text{fictitious}} = 1/T_{\text{fictitious}}$ is the inverse of the fictitious temperature. Comparing against the \renyi entropy $S_n$ in Eq.~\eqref{renyi entropy definition}, we find a proportionality between the \renyi entropies and the thermodynamic free energy of the \entham (thereby connecting the two) when the inverse fictitious temperature $\beta_{\text{fictitious}} = n$
\begin{equation}
    S_{n} = \frac{n}{n-1} \Ff_E(n).
\label{eq:label_connecting_entropy_and_free_energy}
\end{equation}
This concludes our discussion that the fictitious temperature of the \entham is given by $1/n$.

\section{Additional plots for different interacting strengths}
\label{appendix}

\begin{figure}
\centering
%\captionsetup{justification=centering}
\includegraphics[width=0.49\textwidth]{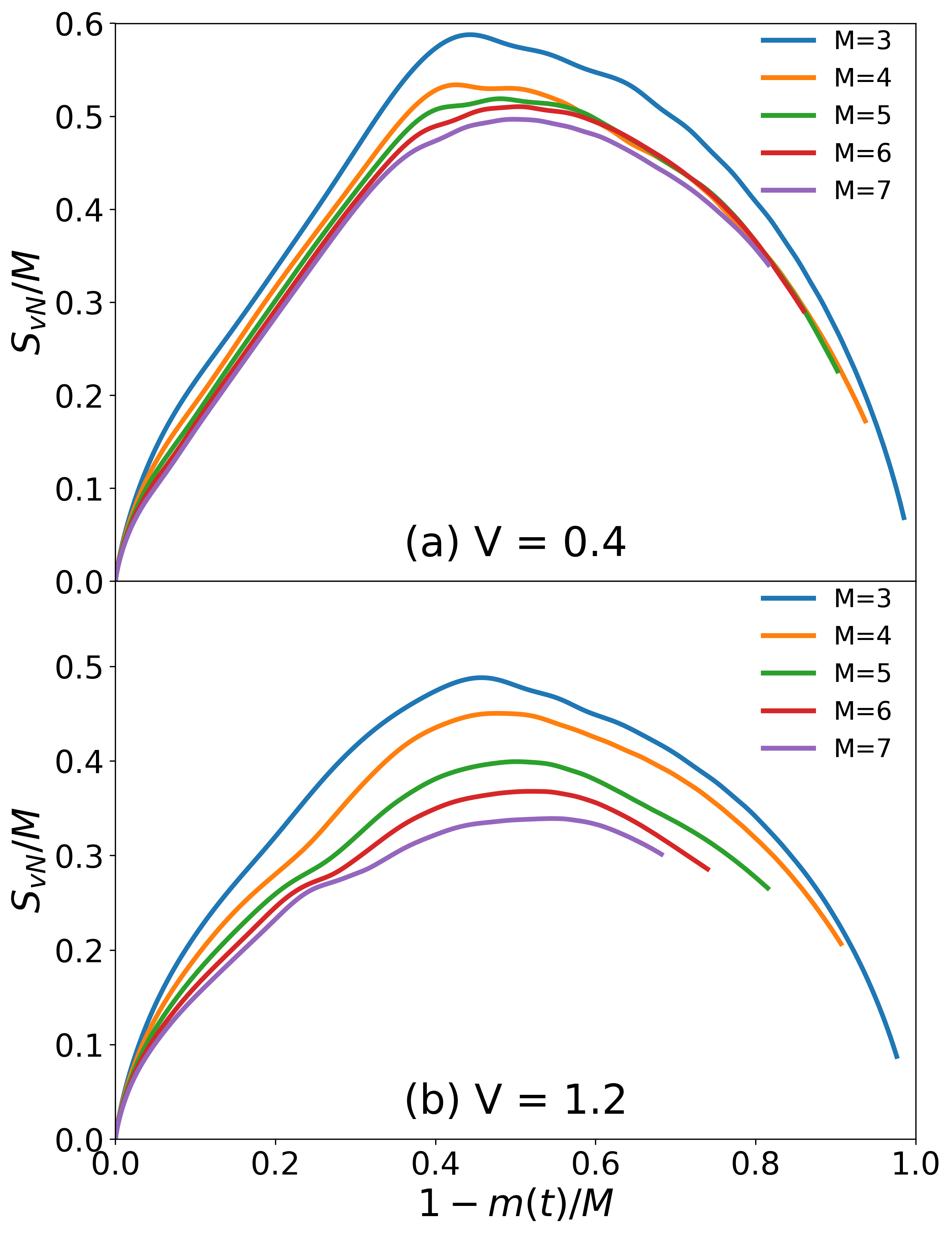}
    \caption{Von Neumann bipartite entanglement entropy density between the system and the environment is plotted against the decayed fraction of particles from the system into the environment. We obtain a Page curve dynamics where the vNEE bends down towards zero. Total number of sites are fixed at $L=50$. The parameters in Eq.~\eqref{eq:model} are fixed at $t_s = t_e = 1.0$ and $g=0.5$. The simulation is run for time starting at $t=0.0$ to a maximum time of $t_{\text{max}} = 40.0$ with time steps $\mathrm{d}t = 0.02$. Plot (a) is for the interaction strength $V=0.4$ while (b) is for the interaction strength $V=1.2$.}
    \label{fig:appendix_ee}
\end{figure}

\begin{figure}
\centering
%\captionsetup{justification=centering}
\includegraphics[width=0.49\textwidth]{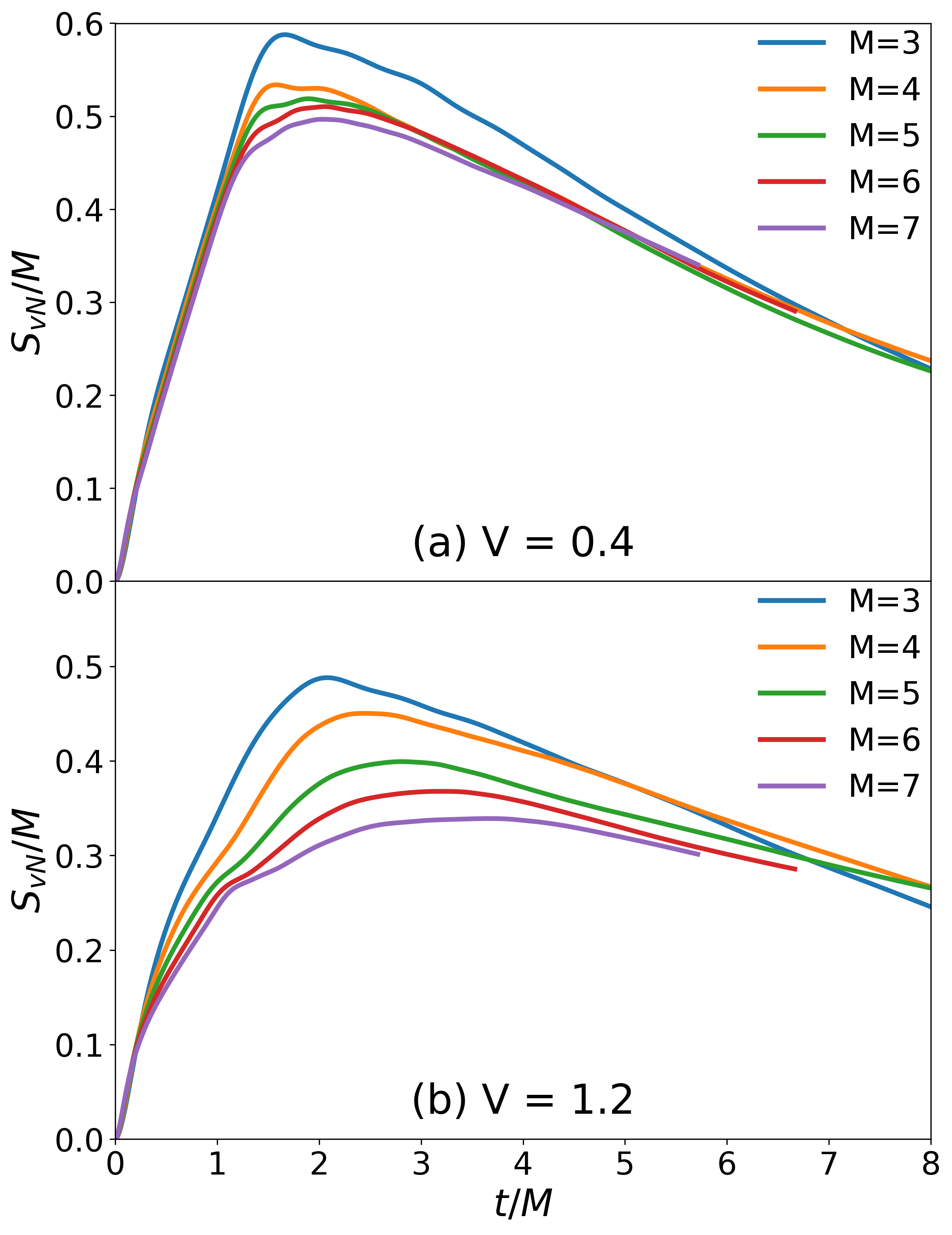}
    \caption{The same behavior as plotted in Fig.~\ref{fig:appendix_ee} where the x-axis is the normalized time instead of decayed fraction of particles from the system into the environment.}
    \label{fig:appendix_ee_in_time}
\end{figure}

We have already provided plots for $V=0.8$ in the main text, here we provide the plots for interaction strengths $V=0.4$ and $V=1.2$ corresponding to different system sizes. There are no qualitative changes and all physics discussed in the main text stay the same, including the sensitivity on the coupling parameter $g$ between the system and the environment for any interacting strength. Fig.~\ref{fig:appendix_ee} shows the behavior of the vNEE for interacting strengths $V=0.4$ and $V=1.2$ as plotted against the decayed fraction of particles from the system into the environment. The same behavior is plotted in Fig.~\ref{fig:appendix_ee_in_time} where the x-axis is changed from decayed fraction of particles to normalized time.

\begin{figure}
\centering
%\captionsetup{justification=centering}
\includegraphics[width=0.49\textwidth]{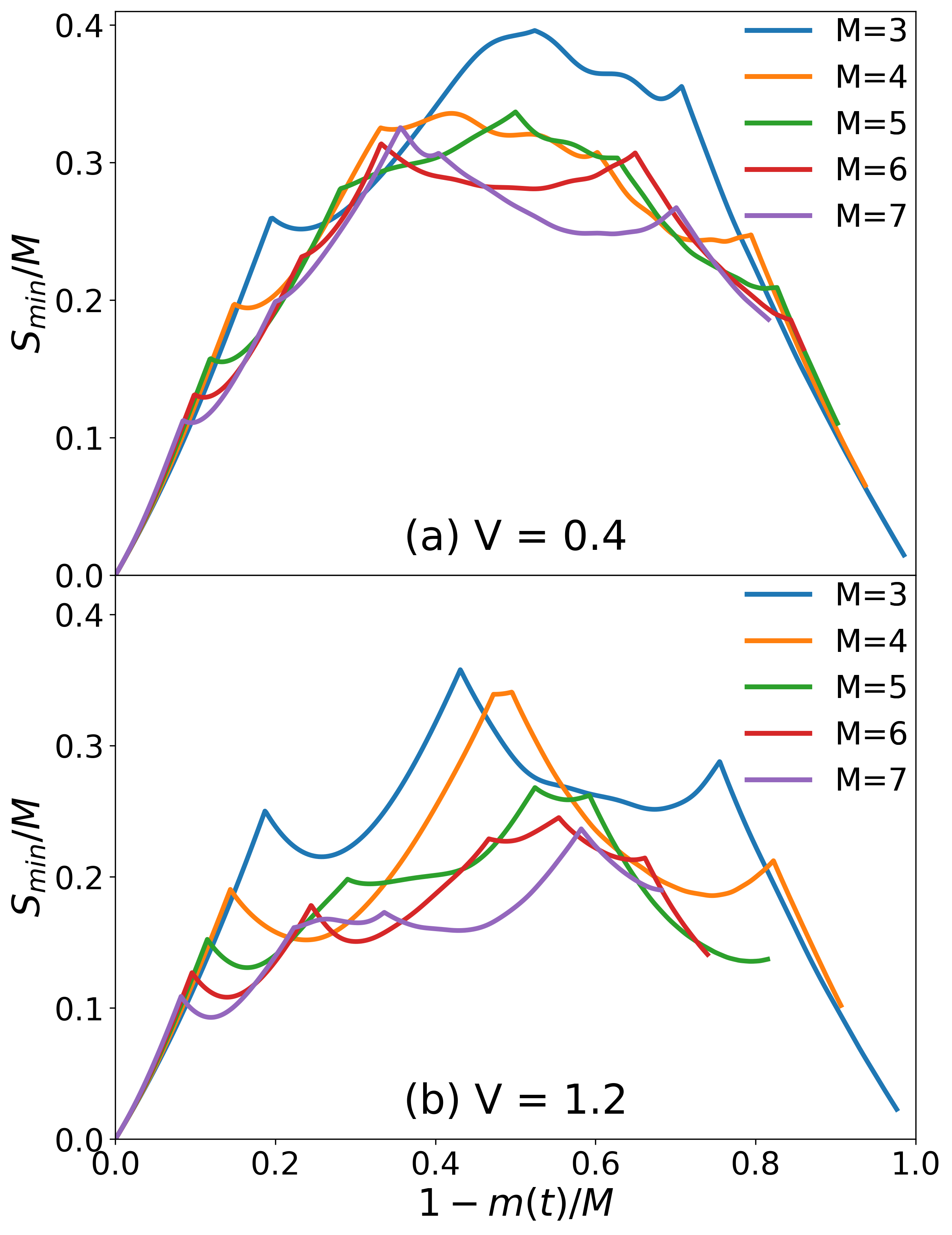}
    \caption{Min-entropy density is plotted against the decayed fraction of particles from the system into the environment. We obtain a Page curve dynamics where the vNEE bends down towards zero. We have used the same parameter values as used in Fig.~\ref{fig:appendix_ee} to plot the vNEE. Plot (a) is for the interaction strength $V=0.4$ while (b) is for the interaction strength $V=1.2$. As motivated by the free case \cite{Kehrein2024Jun}, later non-analyticities gets smeared out in the thermodynamic limit.}
    \label{fig:appendix_min_ent}
\end{figure}

\begin{figure}
\centering
%\captionsetup{justification=centering}
\includegraphics[width=0.49\textwidth]{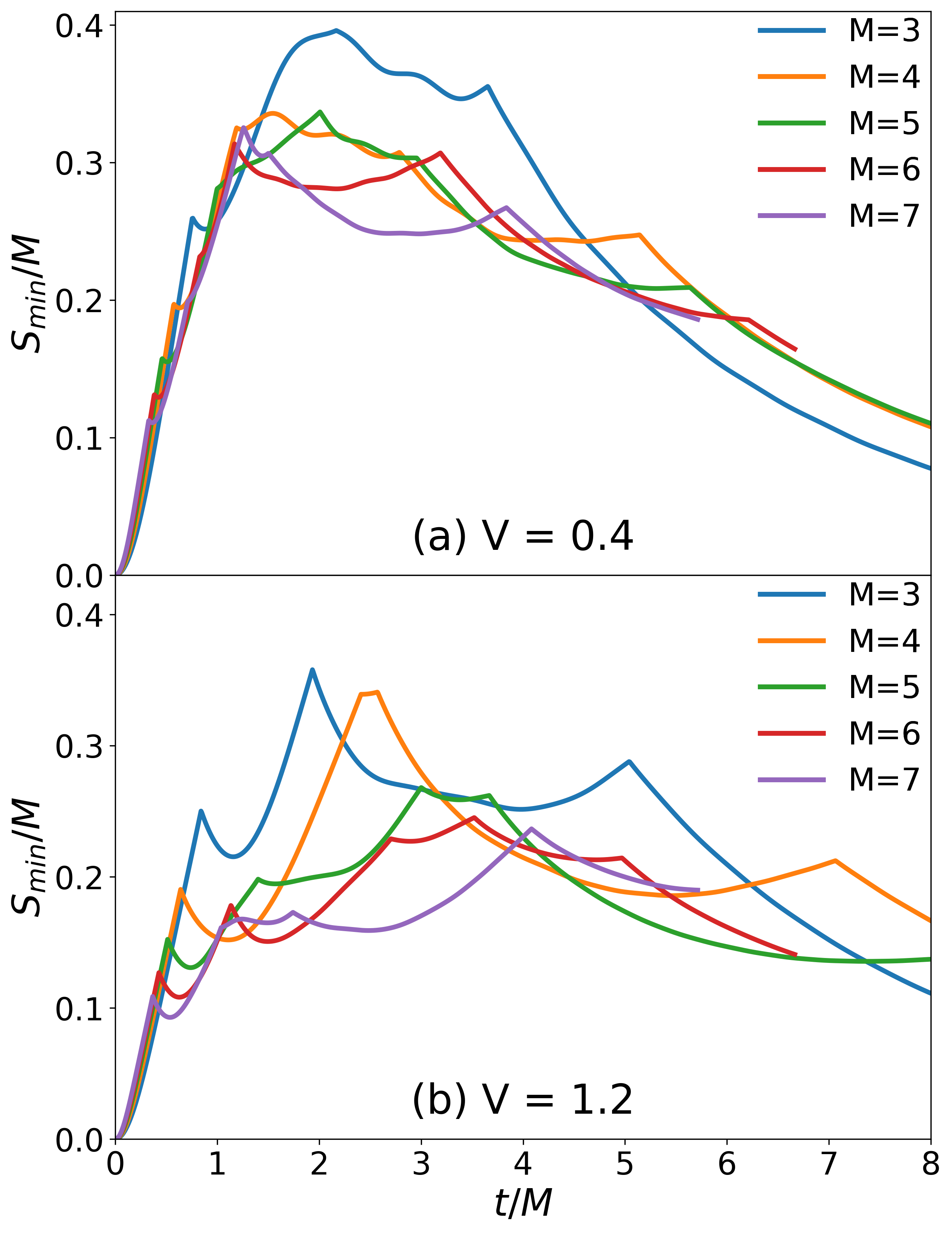}
    \caption{The same behavior as plotted in Fig.~\ref{fig:appendix_min_ent} where the x-axis is the normalized time instead of decayed fraction of particles from the system into the environment.}
    \label{fig:appendix_min_ent_in_time}
\end{figure}

Next we plot the min-entropy in Fig.~\ref{fig:appendix_min_ent} corresponding to the same parameter values as used in Fig.~\ref{fig:appendix_ee} where we again find non-analyticities develop in time. We have also provided the plot in Fig.~\ref{fig:appendix_min_ent_in_time} where instead of the decayed fraction, we have plotted the min-entropy for the normalized time.

\begin{figure}
\centering
%\captionsetup{justification=centering}
\includegraphics[width=0.49\textwidth]{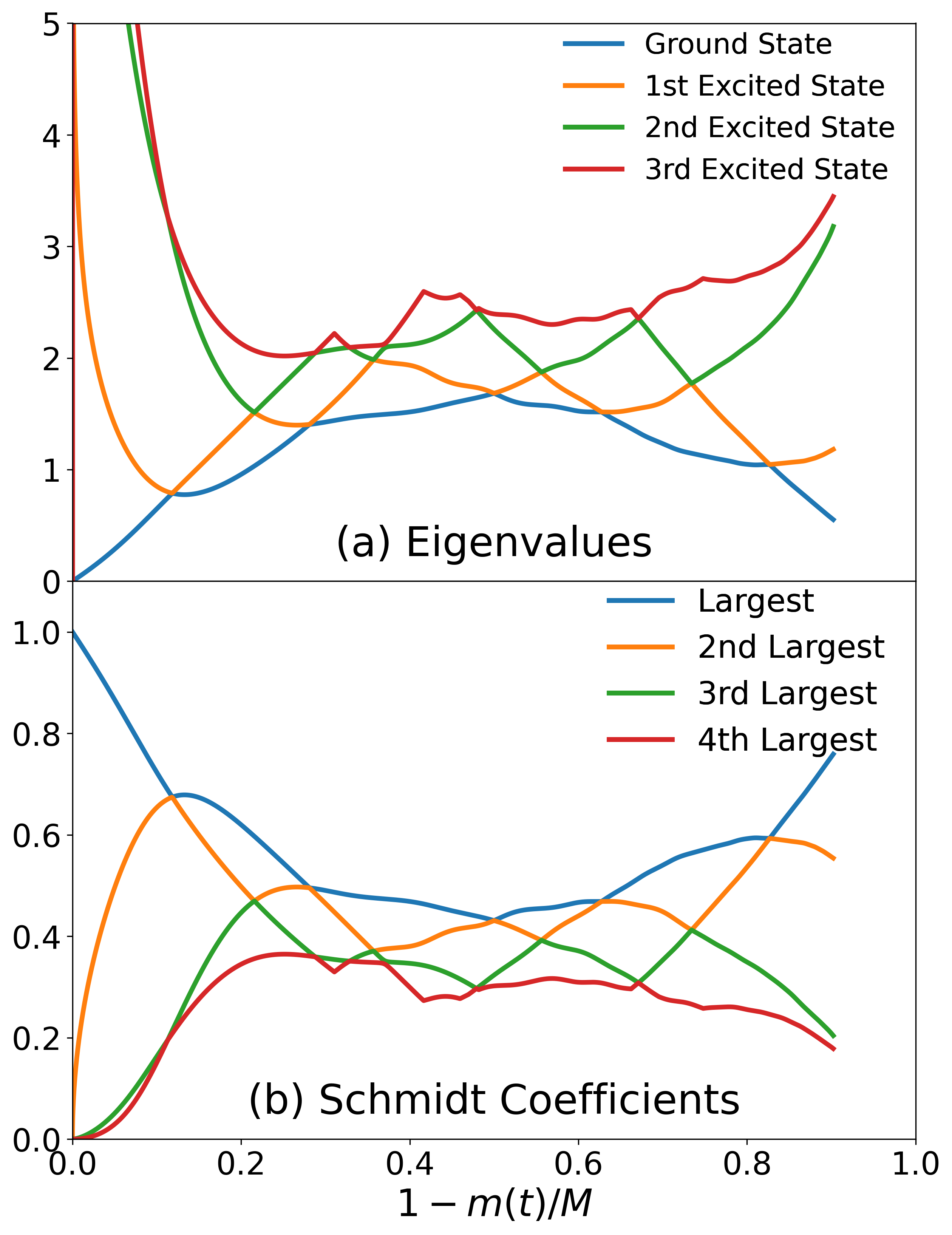}
    \caption{We plot the crossing between the ground state and the first excited state observed in the \entham signifying a quantum phase transition that results in non-analyticity in min-entropy in Fig.~\ref{fig:appendix_min_ent} (or equivalently, Fig.~\ref{fig:appendix_min_ent_in_time}) The parameter values in Eq.~\eqref{eq:model} are: $M=5$, $V=0.4$, $t_s = t_e =1 $, $g=0.5$. Simulation time is from $t=0.0$ to a maximum time of $t_{\text{max}} = 40.0$ with time steps $\mathrm{d}t=0.02$. Only the (a) Eigenvalues of the \entham is plotted against decayed fraction of particles into the environment where quantum phase transition happens at the same time where min-entropy develops non-analyticity. (b) Largest four Schmidt coefficients are plotted against the decayed fraction of particles into the environment. Eigenvalues are extracted from these Schmidt coefficients, as explained in Appendix \ref{subsec. entanglement hamiltonian} (Eq.~\eqref{eq:eigenvalues of entham}). For our purposes, only the ground state (corresponding to the largest Schmidt coefficient) and the first excited state (corresponding to the second-largest Schmidt coefficient) are relevant.}
    \label{fig:appendix_entham_V0.4}
\end{figure}

\begin{figure}
\centering
%\captionsetup{justification=centering}
\includegraphics[width=0.49\textwidth]{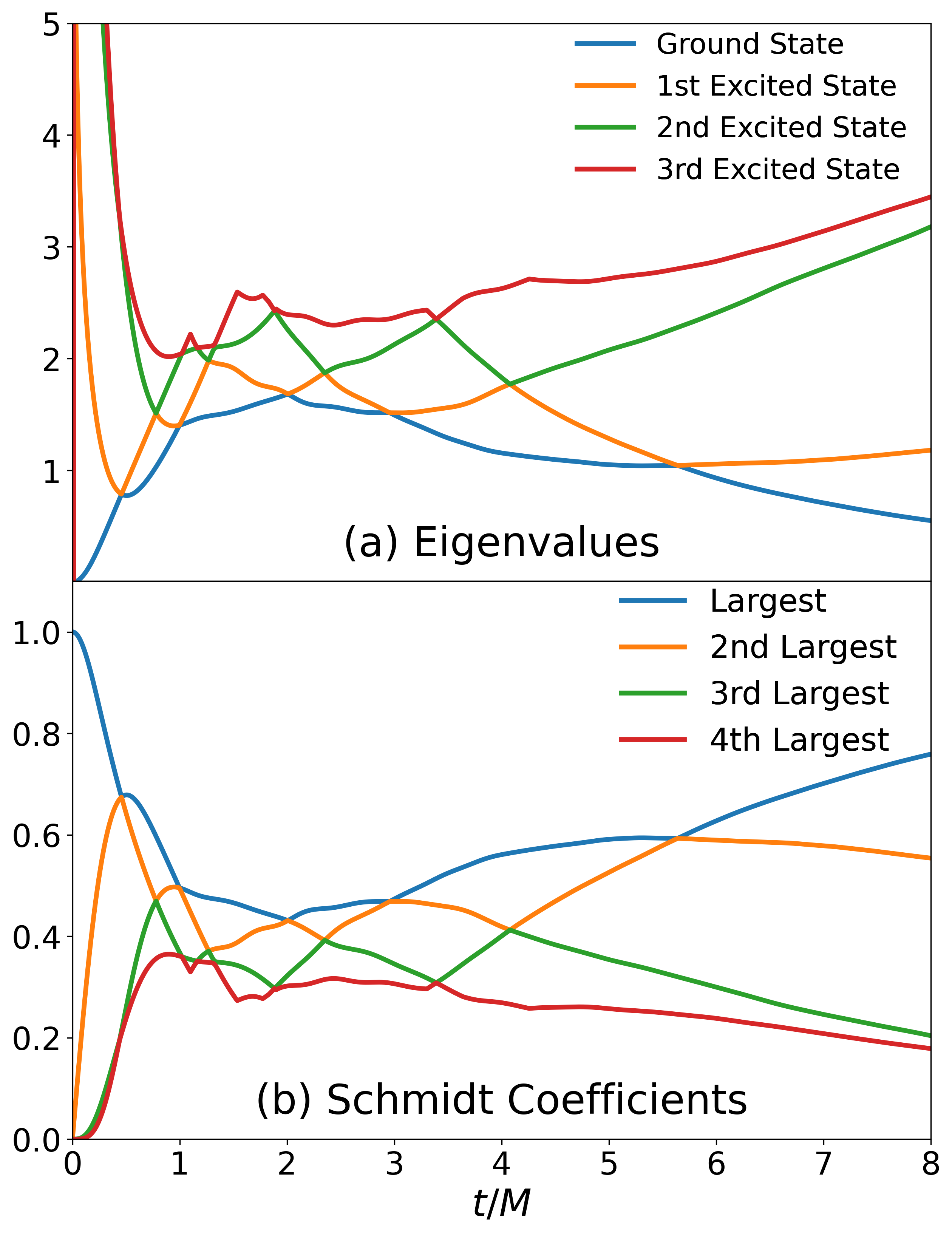}
    \caption{The same behavior as plotted in Fig.~\ref{fig:appendix_entham_V0.4} where the x-axis is the normalized time instead of decayed fraction of particles from the system into the environment.}
    \label{fig:appendix_entham_V0.4_time}
\end{figure}

\begin{figure}
\centering
%\captionsetup{justification=centering}
\includegraphics[width=0.49\textwidth]{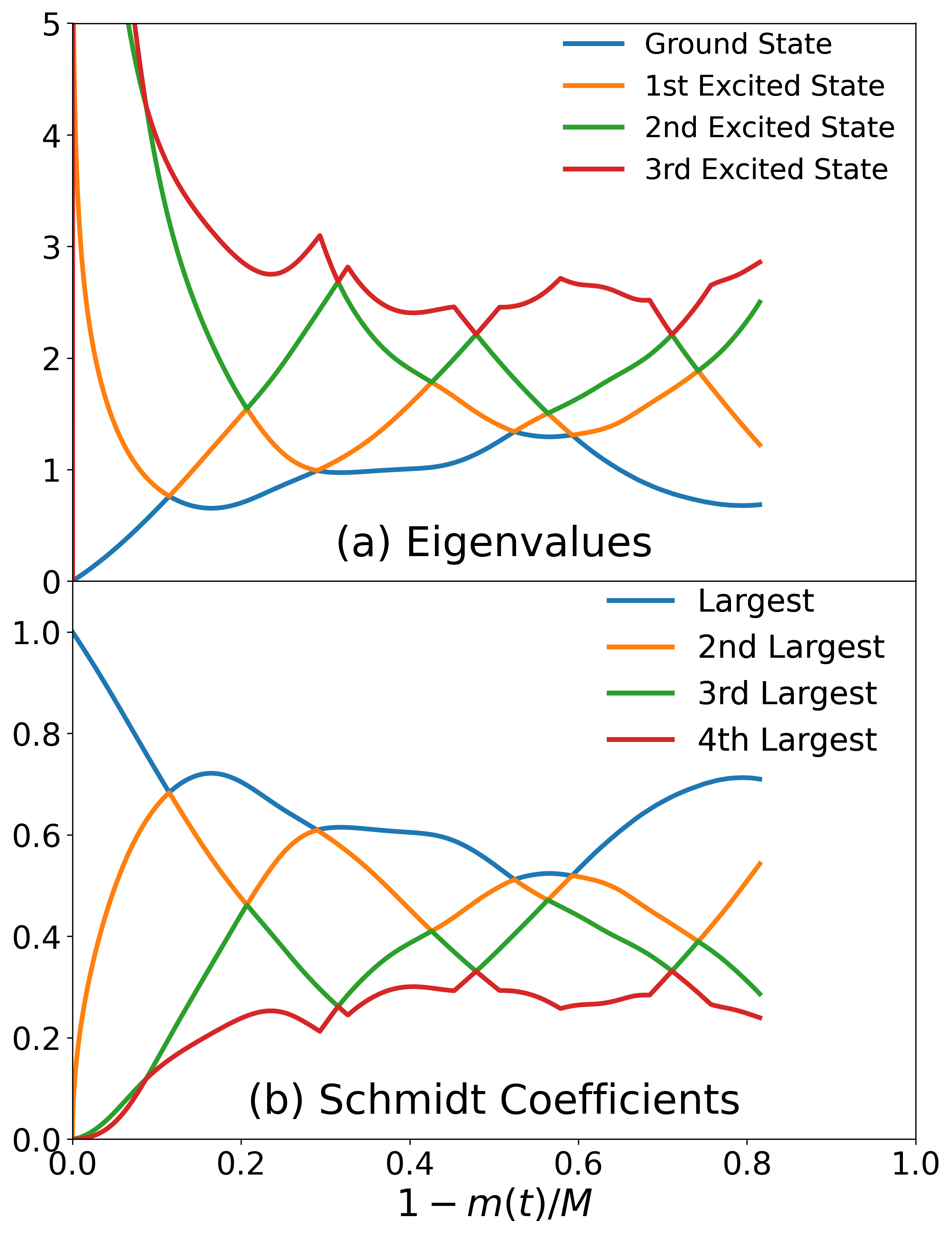}
    \caption{We plot the quantum phase transition that results in non-analyticity in min-entropy for $V=1.2$. The rest of the parameters as well as the conclusions are the same as stated in the captions of Figs.~\ref{fig:entham} and \ref{fig:appendix_entham_V0.4}.}
    \label{fig:appendix_entham_V1.2}
\end{figure}

\begin{figure}
\centering
%\captionsetup{justification=centering}
\includegraphics[width=0.49\textwidth]{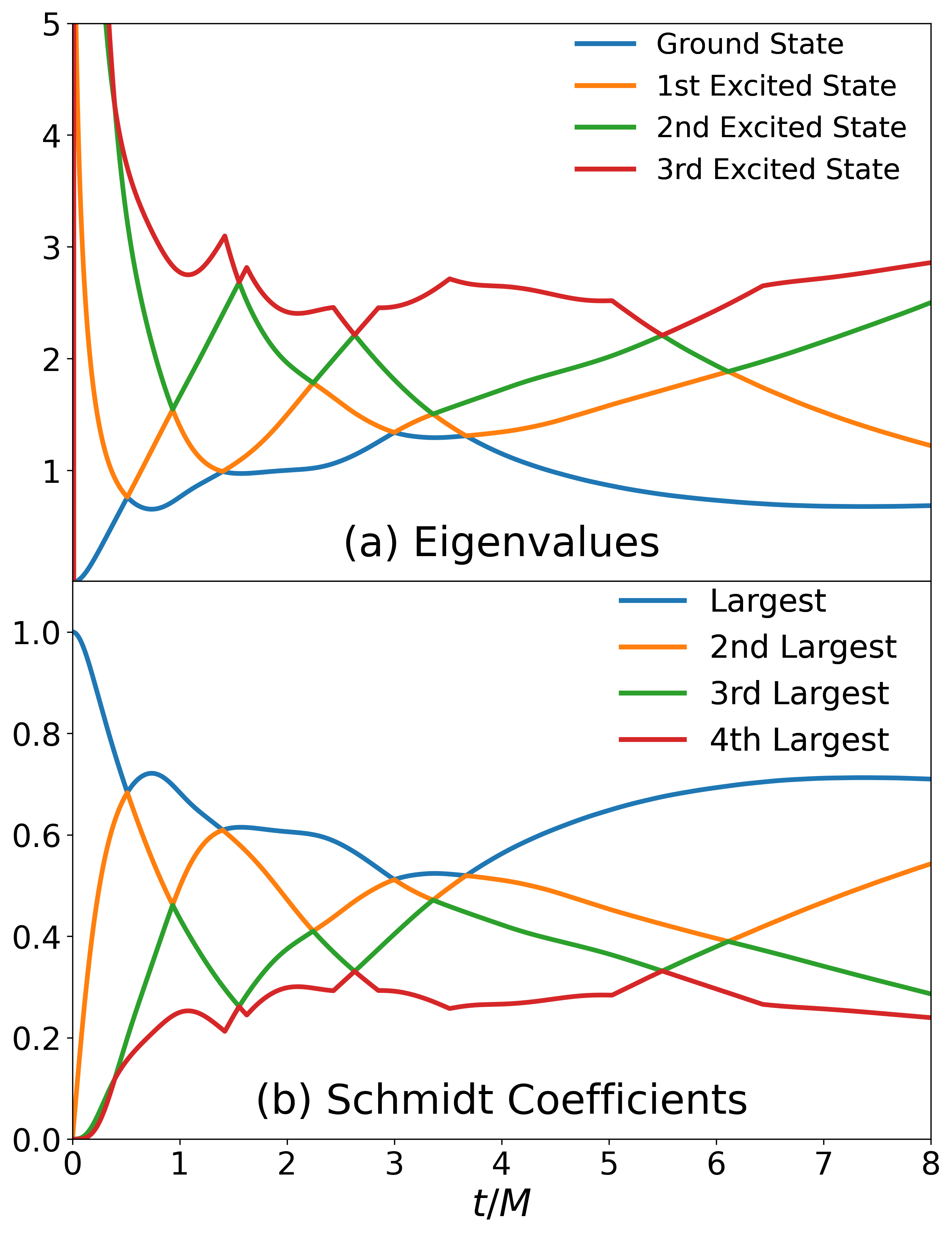}
    \caption{The same behavior as plotted in Fig.~\ref{fig:appendix_entham_V1.2} where the x-axis is the normalized time instead of decayed fraction of particles from the system into the environment.}
    \label{fig:appendix_entham_V1.2_time}
\end{figure}

Next we analyze the entanglement Hamiltonian for these two interacting strengths where we again find an underlying quantum phase transition. We plot the eigenvalues and the Schmidt coefficients (using which the eigenvalues of the entanglement Hamiltonian are calculated) for V= 0.4 (V= 1.2) in Fig.~\ref{fig:appendix_entham_V0.4} (\ref{fig:appendix_entham_V1.2}) against the decayed fraction of particles into the environment, as well as against normalized time in Figs.~\ref{fig:appendix_entham_V0.4_time} and \ref{fig:appendix_entham_V1.2_time}.

\begin{figure}
\centering
%\captionsetup{justification=centering}
\includegraphics[width=0.49\textwidth]{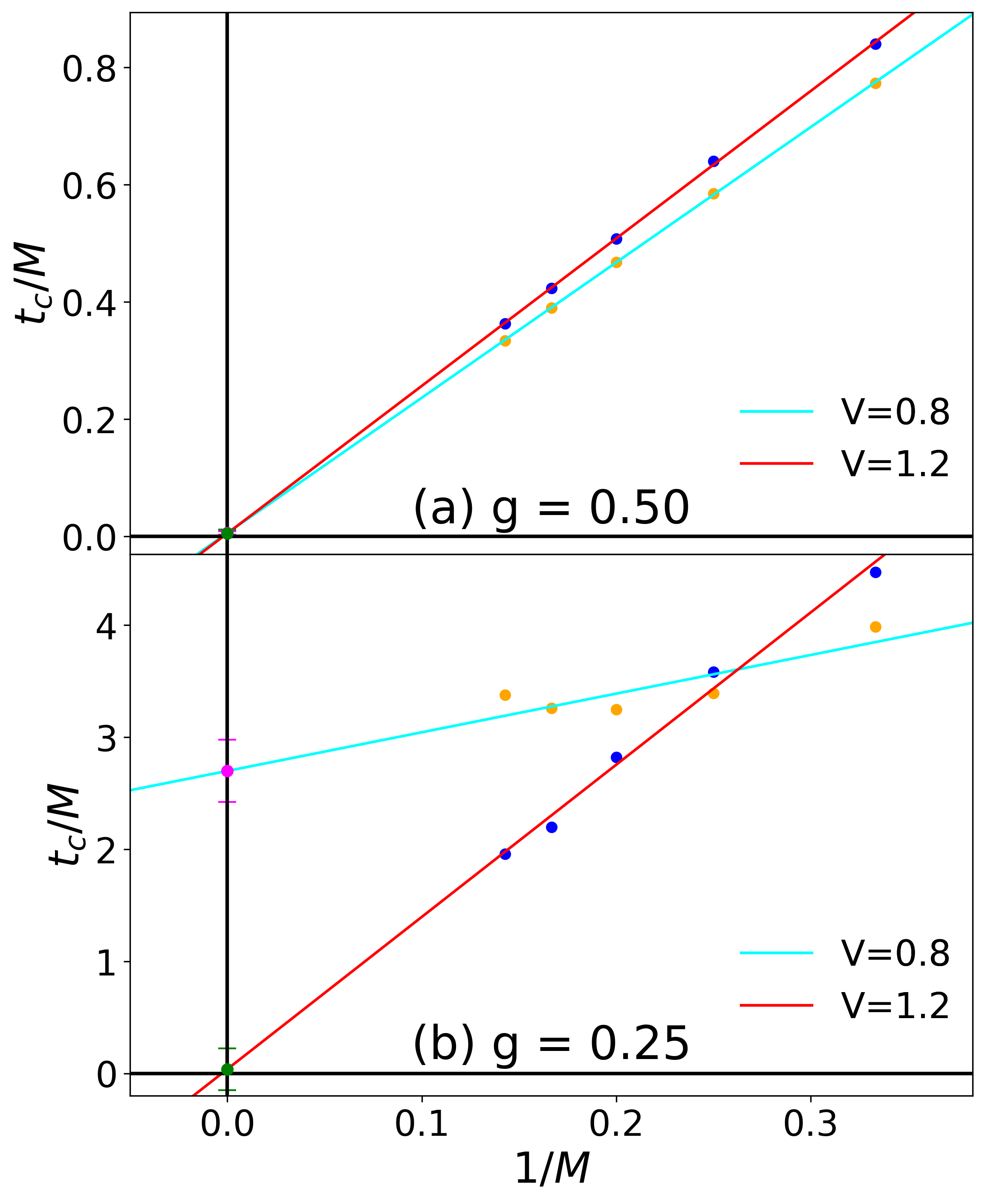}
    \caption{Linear regression plot where we extrapolate the critical normalized time $t_c/M$ to the thermodynamic limit where $M \to \infty$. The system sizes are varied from $M=3$ to $M=7$ for two interaction strengths (total of $5$ data points each): $V=0.8$ and $V=1.2$. Other parameters are $t_s=t_e=1.0$ where the coupling $g$ between the system and the environment is shown in the plot. The regression plot for the same parameter values but for critical decayed fraction (instead of critical time) is given in Fig.~\ref{fig:regression plot}. We find that the critical decayed fraction vanishes in the thermodynamic limit for the stronger coupling $g=0.5$ between the system and the environment, however for weaker coupling $g=0.25$, there exists a non-zero value in the thermodynamic limit. We find an acute sensitivity to the coupling strength which is summarized in Table  \ref{table:appendix_regression_critical_decay_fraction} where even for the weaker coupling, the thermodynamic critical value vanishes with increasing interaction strength $V$.}
    \label{fig:appendix_regression plot}
\end{figure}

We finally perform the regression analysis for normalized time (in contrast to the decayed fraction as done in the main text) corresponding to the first non-analyticity observed in min-entropy. Fig.~\ref{fig:regression plot} provides the regression plot for $V=0.8$ where the thermodynamic extrapolation of critical decayed fraction is done. In Fig.~\ref{fig:appendix_regression plot}, we do a regression analysis for critical normalized time instead and find that it vanishes in the thermodynamic limit. We perform such analyses for a wide-ranging value of interaction strengths to find the extrapolated thermodynamic value of critical time (similar to Table \ref{table:regression_critical_decay_fraction} where we extrapolated to find the critical decayed fraction in the thermodynamic limit) and summarize it in Table \ref{table:appendix_regression_critical_decay_fraction}. A CSV file containing a summary of regression results is provided in the research data management \cite{zenodo}.

\begin{table}
    \centering
    \caption{Linear regression for the first non-analytic points observed in min-entropy (see Figs.~\ref{fig:appendix_min_ent_in_time}) observed across different system sizes ($M=3$ to $M=7$ for a fixed total number of lattice sites $L=50$) that corresponds to the quantum phase transition in the \entham (crossing shown in Figs.~\ref{fig:appendix_entham_V0.4_time} and \ref{fig:appendix_entham_V1.2_time}). In the regression, the $Y-$axis is $t_c/M$ where $t_c$ is the critical time at which the non-analyticity develops in min-entropy. The $X-$axis is $1/M$, therefore the $Y-$intercept is the extrapolation of the critical time in the thermodynamic limit. For intermediate to strong interactions, the thermodynamic value for critical time vanishes, and we provide an analysis in the text based on Fig.~\ref{brick-wall}. We also note an acute sensitivity on the tunneling strength $g$ where the thermodynamic value vanishes at intermediate interaction strength for larger values of $g$.}
    \label{table:appendix_regression_critical_decay_fraction}
    \begin{tabular}{|c|c|c|c|c|c|c|c|}
        \hline
        V & g & $Y-$intercept & Slope & $R^2$ \\
        \hline
        \multirow{2}{*}{$0.0$} & $0.50$ & $-0.0005 \pm 0.0038 $ & $2.2669 \pm 0.0166$& $0.9998 $\\  
         & $0.25$ & $2.3382 \pm 0.4340 $& $4.7963 \pm 1.8969 $& $0.6806 $\\ 
        \hline
        \multirow{2}{*}{$0.4$} & $0.50$ & $0.0060 \pm 0.0028 $ & $2.2466 \pm 0.0121 $& $0.9999 $\\
        & $0.25$ & $2.4204 \pm 0.0684 $& $4.4464 \pm 0.2988 $& $0.9866 $\\ 
        \hline
        \multirow{2}{*}{$0.8$} & $0.50$ & $0.0060 \pm 0.0028 $ & $2.3066 \pm 0.0121   $& $0.9999  $\\
        & $0.25$ & $2.6983 \pm 0.2782 $& $3.4475 \pm 1.2160 $& $0.7282   $\\ 
        \hline
        \multirow{2}{*}{$1.2$} & $0.50$ & $0.0055 \pm 0.0063 $ & $2.5135 \pm 0.0277 $& $0.9996  $\\
        & $0.25$ & $0.0379 \pm 0.1865  $& $13.5767 \pm 0.8150 $& $0.9893 $\\ 
        \hline
    \end{tabular}
\end{table}

\section{Strongly Interacting Case}
\label{app. strongly interacting case}
\begin{figure*}
    \centering % <-- added
 \begin{subfigure}{0.49\textwidth}\hfill
 %\caption{}\label{fig:V=3_Ent}
  \includegraphics[width=\linewidth]{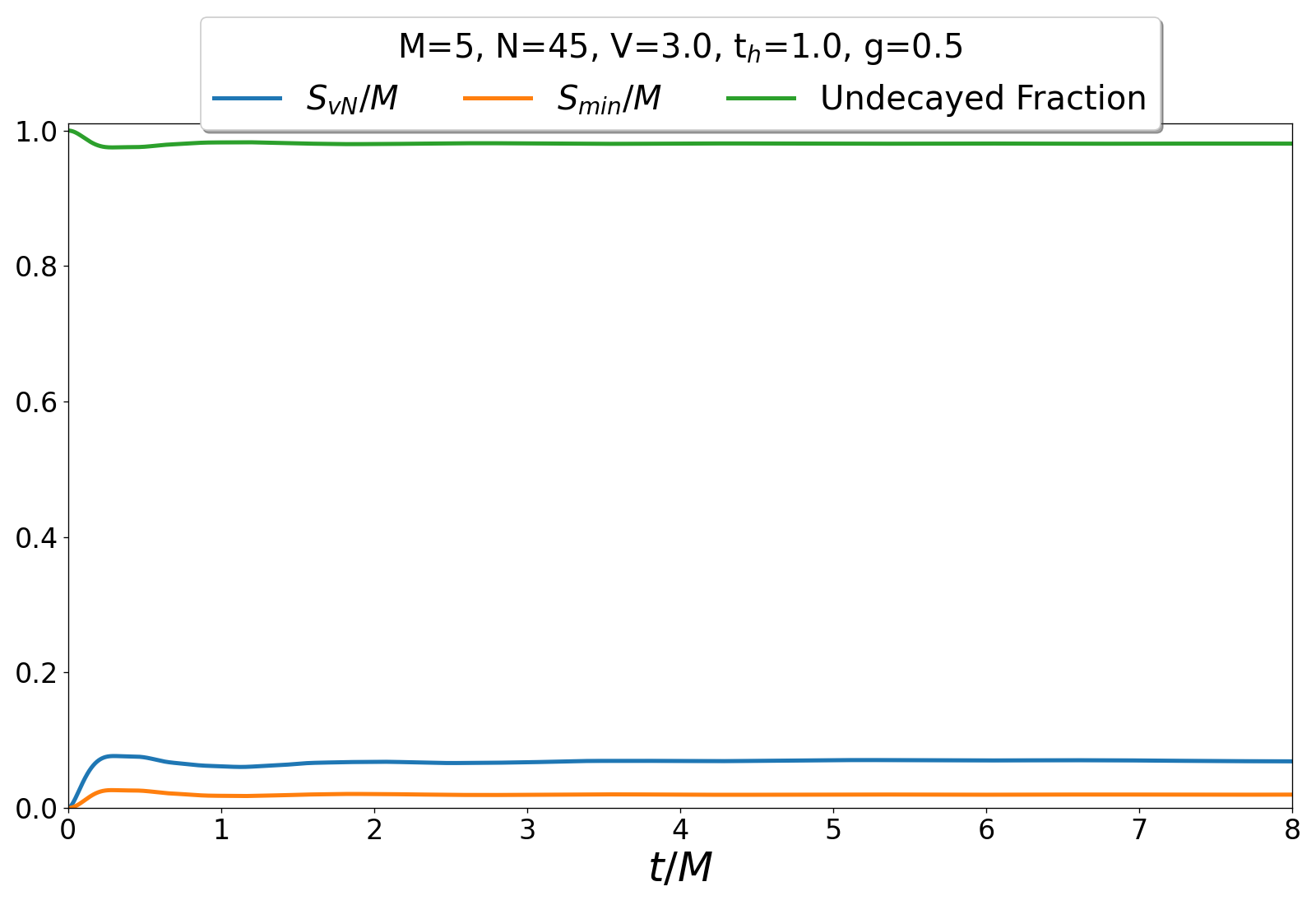}
\end{subfigure} % 
\begin{subfigure}{0.49\textwidth}\hfill
  %\caption{}\label{fig:V=5_Ent}
  \includegraphics[width=\linewidth]{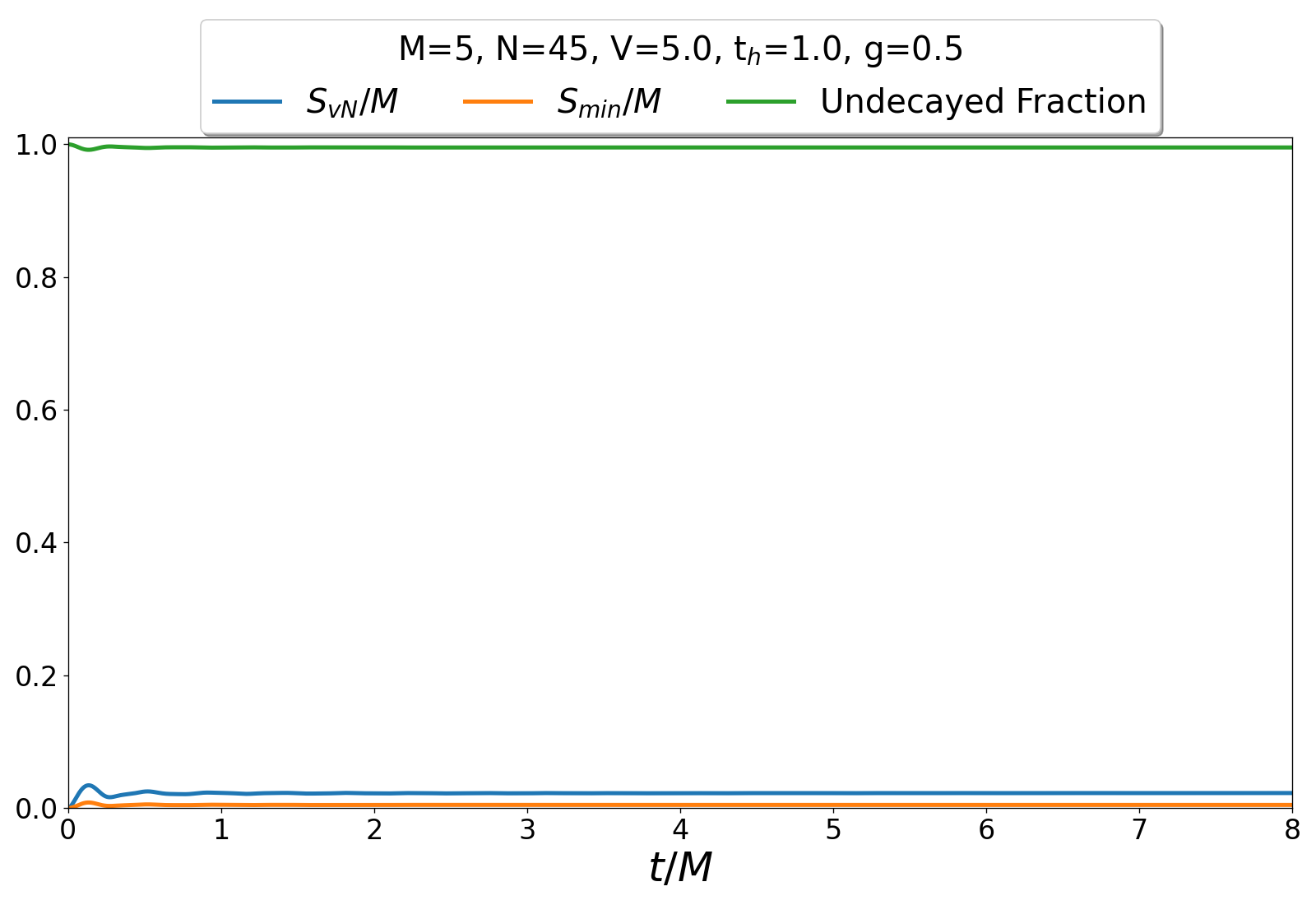}
\end{subfigure}% 
\caption{Results for strongly interacting cases for (a) V=3.0 and (b) V=5.0 where other parameter values are mentioned in the plot. As explained in the main text in Section \ref{sec. regression}, the system gets stuck despite showing short term fluctuations.}
\label{fig:Entropies in strongly interacting case}
\end{figure*}

\begin{figure*}
    \centering % <-- added
 \begin{subfigure}{0.49\textwidth}\hfill
 %\caption{}\label{fig:V=3_Schmidt}
  \includegraphics[width=\linewidth]{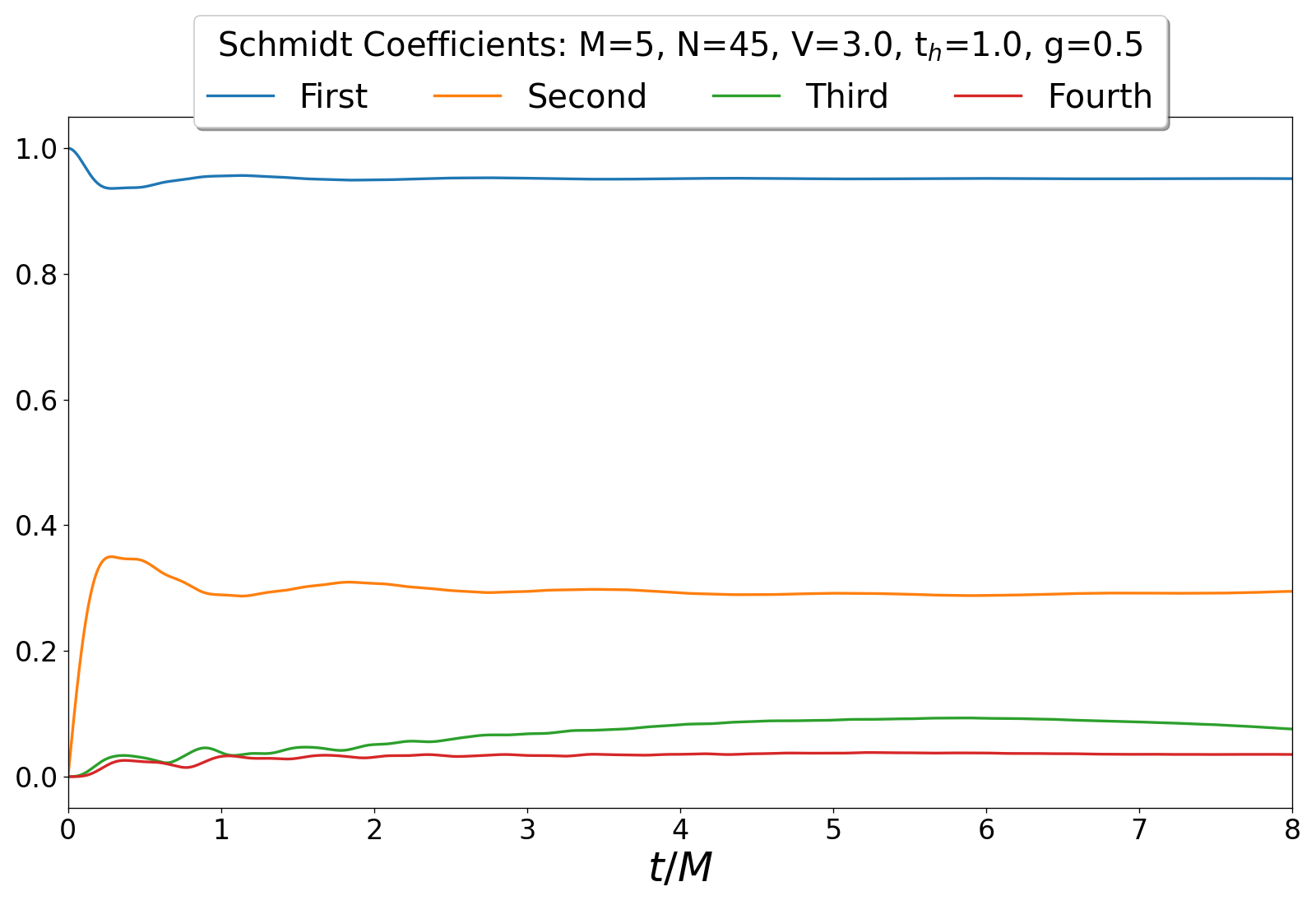}
\end{subfigure} % 
\begin{subfigure}{0.49\textwidth}\hfill
  %\caption{}\label{fig:V=5_Schmidt}
  \includegraphics[width=\linewidth]{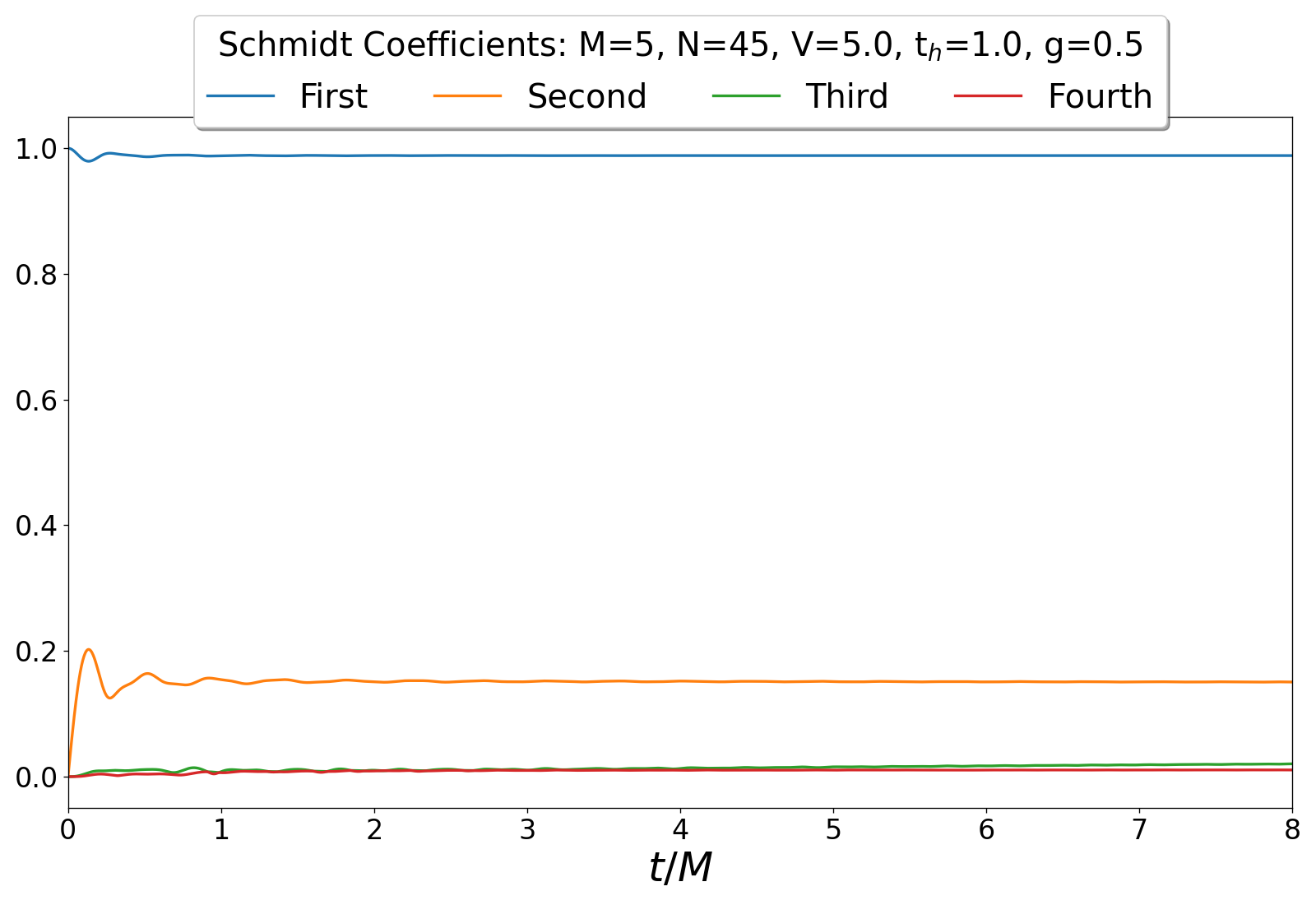}
\end{subfigure}% 
\caption{The corresponding Schmidt coefficients for the graphs shown in Fig.~\ref{fig:Entropies in strongly interacting case} for strongly interacting cases for (a) V=3.0 and (b) V=5.0 where other parameter values are mentioned in the plot. As explained in the main text in Section \ref{sec. regression}, only the largest and the second-largest Schmidt coefficients play a role which becomes evident in these plots as well. This allows us to write the ansatz for the particle number in the system, as in Eq.~\eqref{eq:ansatz}.}
\label{fig:Schmidt in strongly interacting case}
\end{figure*}

As discussed in the analysis in Section \ref{sec. regression} in the main text, in the strongly interacting case where $V/t_h \gg 1$ (where $t_h \equiv t_s = t_e$ which has been set to $1.0$ in this work), a maximum of one particle can leak into the environment. Accordingly, we were able to write down an ansatz for the number of particles in the system (Eq.~\eqref{eq:ansatz}) where only the largest ($1- \lambda(t) $) and the second-largest ($\lambda(t)$) Schmidt coefficients play a role such that $\lambda(t=0) =0$. We show the results for $V=3.0$ and $V=5.0$ in Fig.~\ref{fig:Entropies in strongly interacting case} that indeed the system gets stuck despite showing short time fluctuations and there is no further leakage of particle, at least for the timescale accessible to us. We further plot in Fig.~\ref{fig:Schmidt in strongly interacting case} the four largest Schmidt coefficients where we find (as claimed in the main text) that only the largest and the second-largest Schmidt coefficients play a dominating role in determining the temporal dynamics of the system as the interaction strength is further increased. As mentioned above Eq.~\eqref{eq:ansatz}, strong repulsive force not necessarily lead to particles dispersing, but the dynamics can completely freeze in the presence of strong interactions. This is what we observe here. 

\onecolumngrid

\twocolumngrid

\section{Further Plots for Finite Size Scaling and Critical Exponent}
\label{app. exponent}

\begin{figure}
\centering
%\captionsetup{justification=centering}
\includegraphics[width=0.49\textwidth]{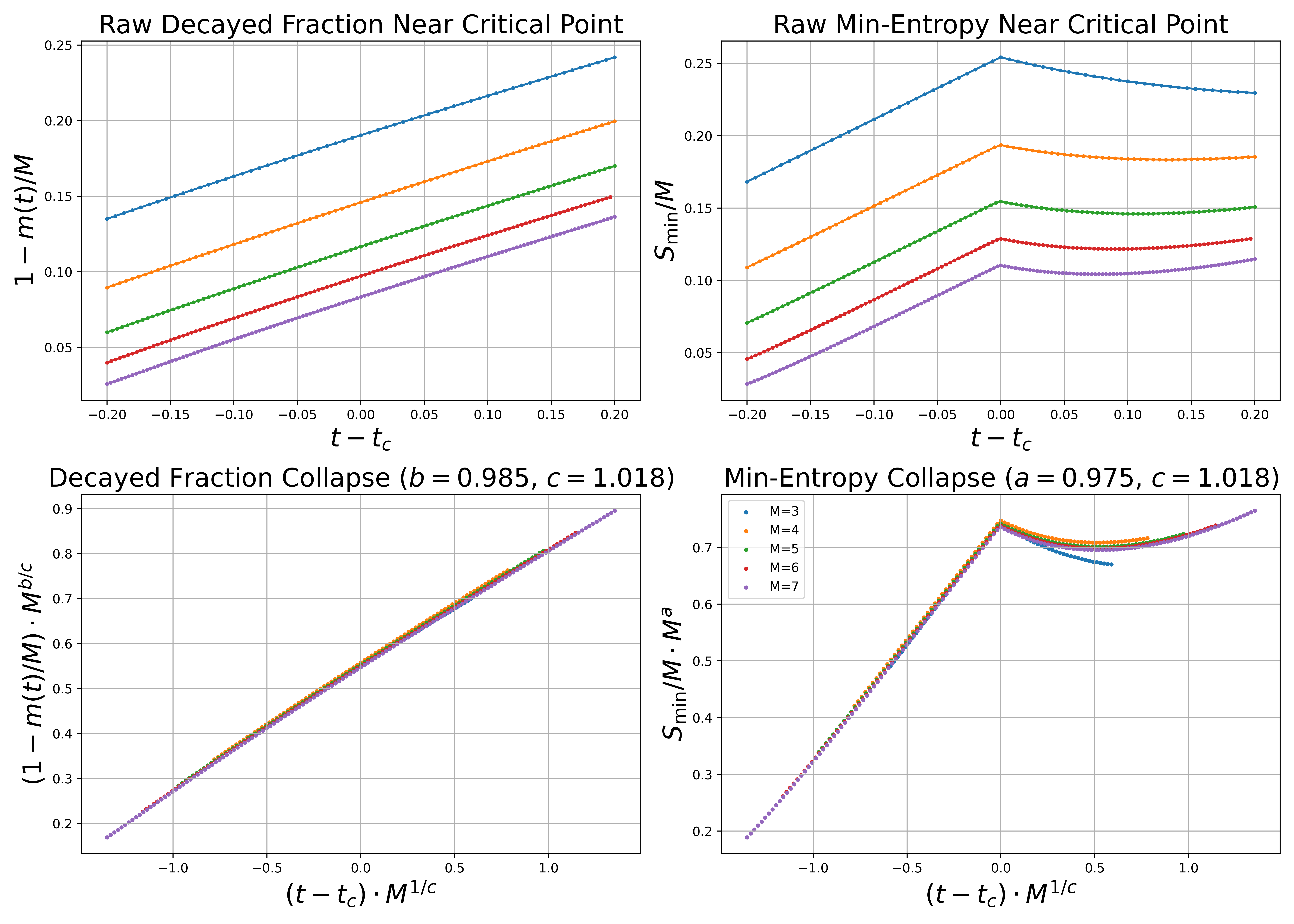}
    \caption{Parameters: $V=0.8$, $g=0.5$, $t_s = t_e = 1.0$, $L=50$ and $M=\{3,4,5,6,7\}$.}
    \label{fig:collapse plot V=0.8}
\end{figure}

\begin{figure}
\centering
%\captionsetup{justification=centering}
\includegraphics[width=0.49\textwidth]{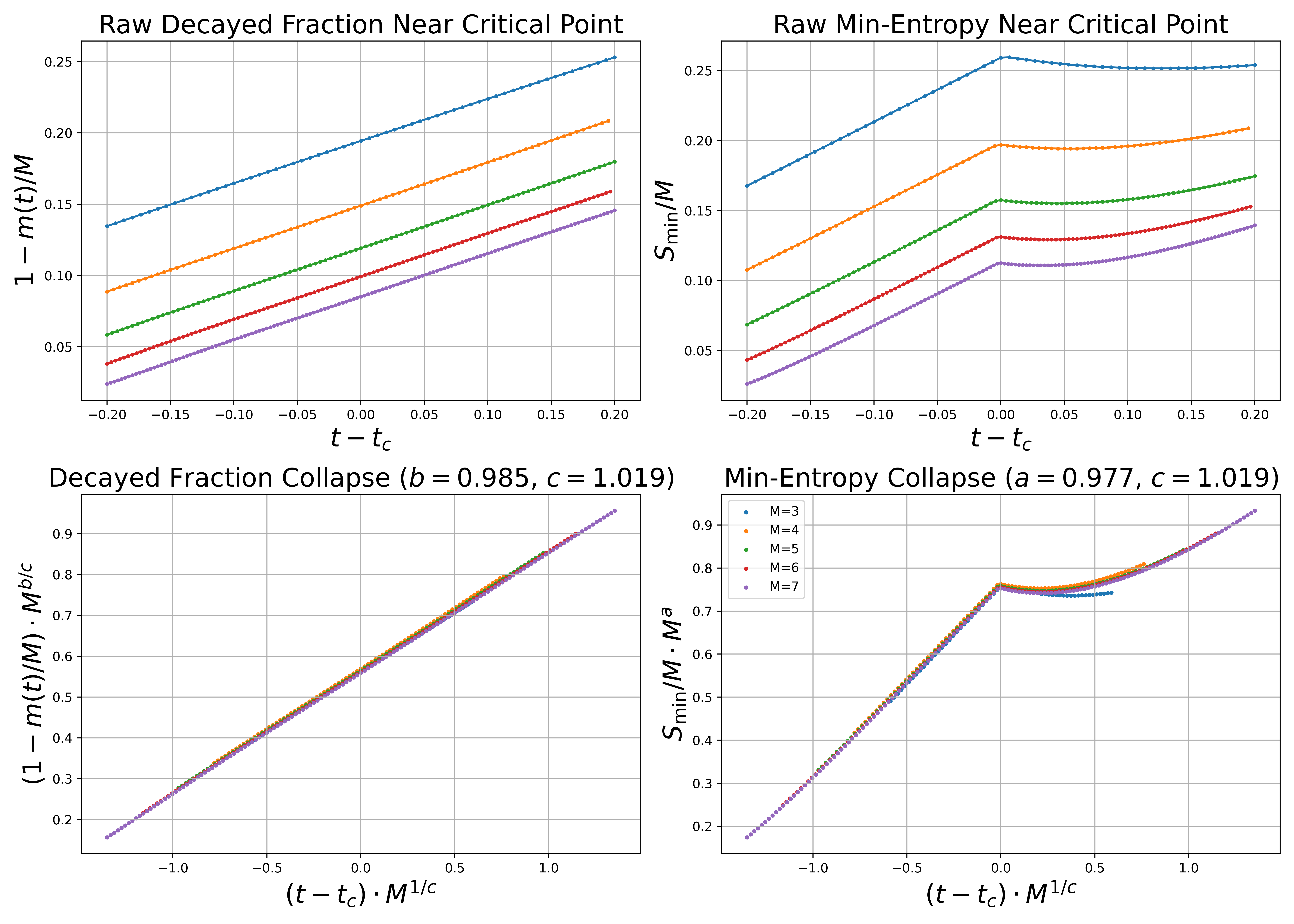}
    \caption{Parameters: $V=0.4$, $g=0.5$, $t_s = t_e = 1.0$, $L=50$ and $M=\{3,4,5,6,7\}$.}
    \label{fig:collapse plot V=0.4}
\end{figure}

\begin{figure}
\centering
%\captionsetup{justification=centering}
\includegraphics[width=0.49\textwidth]{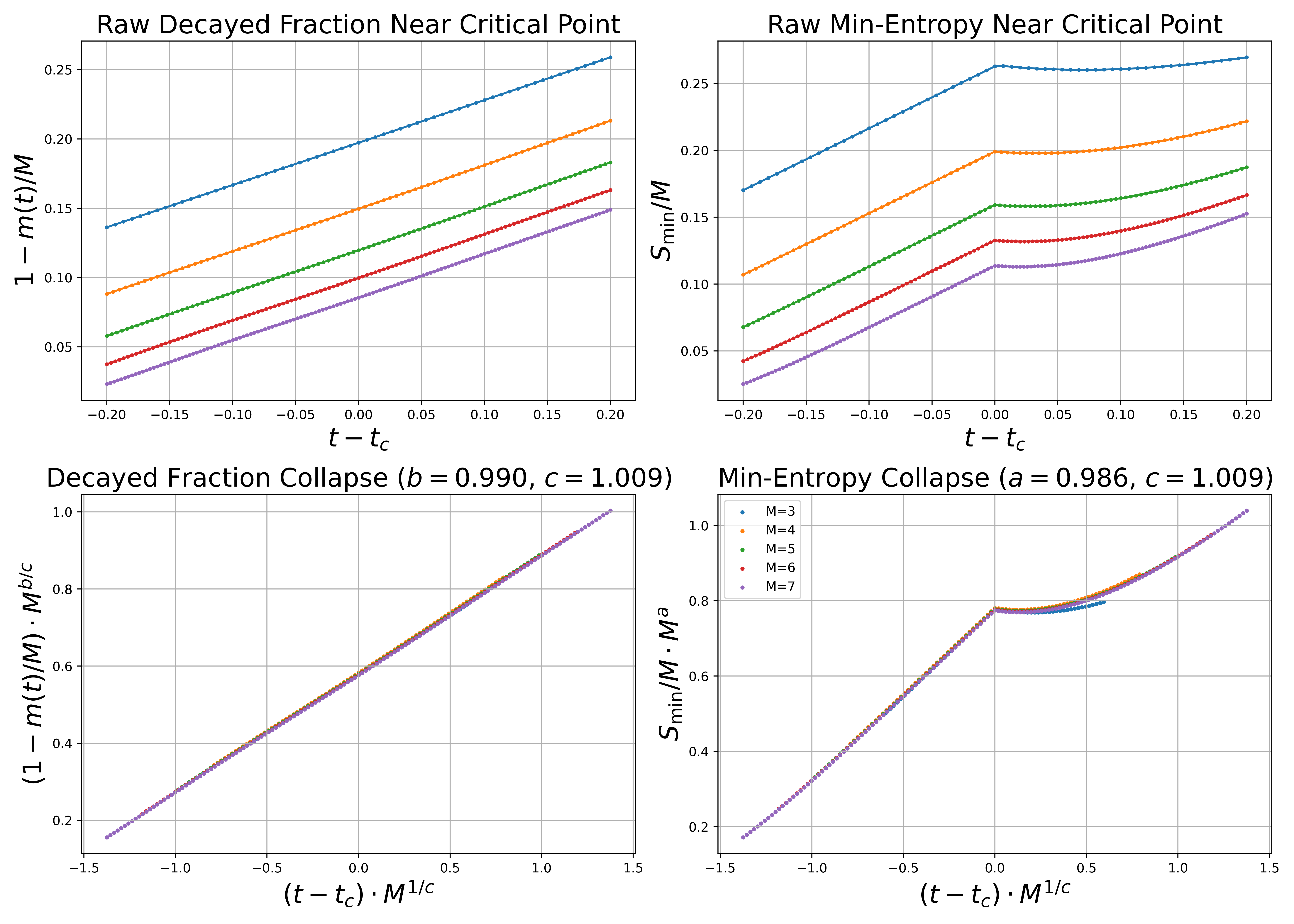}
    \caption{Parameters: $V=0.0$, $g=0.5$, $t_s = t_e = 1.0$, $L=50$ and $M=\{3,4,5,6,7\}$. This is also used for benchmarking as the free case is solvable analytically in the thermodynamic limit \cite{Kehrein2024Jun}.}
    \label{fig:collapse plot V=0.0}
\end{figure}

We present here the plots for $V=0.0$, $V=0.4$ and $V=0.8$ in Figs.~\ref{fig:collapse plot V=0.8}, \ref{fig:collapse plot V=0.4} and \ref{fig:collapse plot V=0.0}. The descriptions of the top and the bottom rows in each of these plots are explained in the caption of Fig. \ref{fig:collapse plot V=1.2}.

\end{document}